\documentclass[11pt,a4paper]{article}
\usepackage{jheppub,amsmath,amsthm,amssymb,slashed,url}
\usepackage{graphicx}
\usepackage[all]{xy}
\font\teneurm=eurm10 \font\seveneurm=eurm7 \font\fiveeurm=eurm5
\newfam\eurmfam
\textfont\eurmfam=\teneurm \scriptfont\eurmfam=\seveneurm
\scriptscriptfont\eurmfam=\fiveeurm

\font\teneusm=eusm10 \font\seveneusm=eusm7 \font\fiveeusm=eusm5
\newfam\eusmfam
\textfont\eusmfam=\teneusm \scriptfont\eusmfam=\seveneusm
\scriptscriptfont\eusmfam=\fiveeusm

\font\tencmmib=cmmib10 \skewchar\tencmmib='177
\font\sevencmmib=cmmib7 \skewchar\sevencmmib='177
\font\fivecmmib=cmmib5 \skewchar\fivecmmib='177
\newfam\cmmibfam
\textfont\cmmibfam=\tencmmib \scriptfont\cmmibfam=\sevencmmib
\scriptscriptfont\cmmibfam=\fivecmmib
\numberwithin{equation}{section}

\newcommand{\dl}{\delta} %delta
\newcommand{\Dl}{\Delta}
\newcommand{\e}{\epsilon} %epsilon
\newcommand{\ve}{\varepsilon}

\newcommand{\g}{\gamma} %gamma

\newcommand{\G}{\Gamma}
\newcommand{\la}{\lambda} %lambda
\newcommand{\bla}{\bar{\lambda}}
\newcommand{\La}{\Lambda}
\newcommand{\om}{\omega} %omega

\newcommand{\Om}{\Omega}

\newcommand{\bphi}{\bar{\phi}} %phi

\newcommand{\bpsi}{\bar{\psi}} %psi

\newcommand{\sg}{\sigma} %sigma

\newcommand{\Sg}{\Sigma}
\newcommand{\vt}{\vartheta} %theta
\newcommand{\bxi}{\bar{\xi}}

\newcommand{\Acal}{\mathcal{A}}
\newcommand{\Bcal}{\mathcal{B}}

\newcommand{\Dcal}{\mathcal{D}}

\newcommand{\Jcal}{\mathcal{J}}

\newcommand{\Lcal}{\mathcal{L}}

\newcommand{\Ncal}{\mathcal{N}}

\newcommand{\Rcal}{\mathcal{R}}
\newcommand{\Scal}{\mathcal{S}}

\newcommand{\Vcal}{\mathcal{V}}
\newcommand{\Wcal}{\mathcal{W}}

\newcommand{\Zcal}{\mathcal{Z}}

\newcommand{\Hbb}{\mathbb{H}}

\newcommand{\Lbb}{\mathbb{L}}

\newcommand{\Pbb}{\mathbb{P}}
\newcommand{\Rbb}{\mathbb{R}}
\newcommand{\Zbb}{\mathbb{Z}}

\newcommand{\non}{\nonumber} %delete equation number
\newcommand{\lp}{\left(}
\newcommand{\rp}{\right)}
\newcommand{\lc}{\left\{}
\newcommand{\rc}{\right\}}
\newcommand{\lb}{\left[}
\newcommand{\rb}{\right]}

\newcommand{\RA}{\Rightarrow}
\newcommand{\de}{{\rm d}}
\newcommand{\p}{\partial}
\newcommand{\w}{\!\wedge\!}

\newcommand{\Tr}{\text{Tr}}
\newcommand{\diag}{{\rm diag}} %diagonal matrix
\newcommand{\dlz}{\delta_{\xi}}
\newcommand{\dlbz}{\delta_{\bar{\xi}}}
\newcommand{\bF}{\bar{F}}
\title{Supersymmetric R\'enyi Entropy in Two Dimensions}

\preprint{\begin{flushright} OU-HET 883 \end{flushright}}

\author{Hironori Mori}

\affiliation{Department of Physics, Graduate School of Science, Osaka University, \\ Toyonaka, Osaka 560-0043, Japan}
\emailAdd{hiromori@het.phys.sci.osaka-u.ac.jp}

\abstract{
We compute the exact partition function on the branched two-sphere by the localization technique. It is found that it does not depend on a branching parameter $q$, which means that supersymmetric R\'enyi entropy defined by utilizing it is equivalent to the usual entanglement entropy. We also provide the interpretation of the conical singularities on the branched sphere as defects sit on the poles of the nonsingular two-sphere.
}


\begin{document}
\maketitle

%%%%%%%%%%%%%%%%%% section 1 %%%%%%%%%%%%%%%%%%%%%%%%%%%%%%%%%
\section{Introduction}
The supersymmetric localization \cite{Pestun:2007rz} has brought great advances in quantum field theories in diverse dimensions and given direct and exact physical understandings to the nontrivial issues. One of applications of this technique is the supersymmetric R\'enyi entropy (SRE) $\Scal_{q}^{\text{susy}}$ defined by \cite{Nishioka:2013haa}
\begin{align} % SRE
\Scal_{q}^{\text{susy}} = \frac{1}{1 - q} \log \left| \frac{\Zcal_{q}}{\lp \Zcal_{1} \rp^{q}} \right|,
\label{sre}
\end{align}
where $\Zcal_{1}$ and $\Zcal_{q}$ is the supersymmetric partition function on the $d$-dimensional round sphere $S^d$ and on the branched sphere $S_{q}^{d}$ which is a $q$-covering space exhibiting conical singuralities, respectively. This is the supersymmetric extension of the R\'enyi entropy usually evaluated by the replica trick. We can quantitatively interpret this definition from the geometrical viewpoint in the conformal field theories as follows: we consider the R\'enyi entropy on the flat space $\Rbb^{1, d - 1}$ with a $q$-covering spherical entangling surface. Then, this geometry can be mapped to $\Rbb \times \Hbb^{d - 1}$, where $\Hbb^{d - 1}$ is a $( d - 1 )$-dimensional hyperbolic space, due to conformal symmetry, and the evaluation of the R\'enyi entropy is translated into that of a thermal partition function on $\Rbb \times \Hbb^{d - 1}$ with a temperature set by $q$ \cite{Hung:2011nu}. Further, we compactify the time direction $\Rbb$ with a radius proportional to $q$ and again can comformally map to the $q$-branched sphere $S_{q}^{d}$. The (supersymmetric) R\'enyi entropy define by the density matrix on the original space may be rewritten in terms of the partition function on $S_{q}^{d}$ based on this geometrical transition. The $q$-covering structure in the original setup is seen as the conical singularities on the poles of the sphere. The SRE was originally discussed in three dimensions \cite{Nishioka:2013haa} and used to do the precision test of AdS$_{4}$/CFT$_{3}$ in \cite{Huang:2014gca, Nishioka:2014mwa}. In this context, the SRE is perfectly dual to the entropy of the charged topological black hole (TBH), hence this type of the duality is called TBH$_{d + 1}$/qSCFT$_{d}$ as which we refer to the superconformal field theory on $S_{q}^{d}$. The extensions of this correspondence to $d = 4$ and $5$ have been nontrivially established in \cite{Huang:2014pda, Crossley:2014oea, Alday:2014fsa, Hama:2014iea}.

The focus of interest in this paper is $d = 2$. We exactly derive the partition function on $S_{q}^{2}$ using the localization as done on $S^2$ \cite{Benini:2012ui, Doroud:2012xw} and the squashed sphere $S_{b}^{2}$ \cite{Gomis:2012wy}. We reveal that the final partition function does not depend on the branching parameter $q$, which is not the case of other dimensions. This fact means that the SRE in two dimensions is nothing but the standard entanglement entropy and is consistent with a single interval case, that is, the two-point function in \cite{Giveon:2015cgs} where the SRE have been computed in terms of the correlation function of supersymmetric twisted chiral fields. In addition, it is known that the SRE can be described in terms of the codimension-2 defects keeping supersymmetry placed on nonsingular geometry as discussed in \cite{Nishioka:2013haa}. In comparison with exact computations with the defects for the dynamical gauge field accomplished in \cite{Hosomichi:2015pia}, we also observe that the defect interpretation can work on the effects from the conical singularities on $S_{q}^2$, which substantiates connection of the defects and geometrical singularities.

In the rest of the paper, we calculate the exact partition function on the branched sphere and argue the SRE in two dimensions in Section \ref{Exact}. In Section \ref{Defect}, we discuss these exact results by comparing with the partition function in the presence of the defects on the sphere. In Section \ref{Discussions}, we comment on some open questions and future works. Convension and the detail of the localization are summarized in Appendix \ref{Pre} and \ref{Detail}. Appendix \ref{Ex22} contains the calculation of correct R-charges in some $\Ncal = ( 2, 2 )$ superconformal theories in order to justify the defect interpretation.

%%%%%%%%%%%%%%%%%% section 2 %%%%%%%%%%%%%%%%%%%%%%%%%%%%%%%%%
\section{Exact results on the branched sphere} \label{Exact}

%%%%%%%%%%%%%%%%%% section 2.1 %%%%%%%%%%%%%%%%%%%%%%%%%%%%%%%%
\subsection{Supersymmetry}

\paragraph{Branched and resolved sphere.} %%%%%%%%%%%%%%%%%%%%%%%%%%%%%%%%%%
We start with the $q$-branched two-sphere $S_{q}^{2}$ whose metric is given by
\begin{align} % metric for branched sphere
\de s_{q}^{2} = \ell^{2} \lp \de \vartheta^{2} + q^{2} \sin^{2} \vartheta \de \tau^{2} \rp,
\label{qmetric}
\end{align}
where $\vartheta \in [ 0, \pi ]$ and $\tau \in [ 0 , 2 \pi ]$. This space has conical singularities on the north and south pole, that is, its scalar curvature $\Rcal_{q}$ exhibits the delta function behavior at both poles:
\begin{align} % curvature
\Rcal_{q}
=
\frac{2}{\ell^{2}} \lb 1 + \frac{\frac{1}{q} - 1}{\sin \vartheta} \lp \dl ( \vartheta ) + \dl ( \pi - \vartheta ) \rp \rb.
\label{qcurvature}
\end{align}
We need to impose suitable boundary conditions on the fields at the singularities, but it is technically hard to treat with them in computing some physical quantities. Instead, we would like to put the theory on the sphere smoothing the conical singularities which we call the resolved two-sphere $S_{\e}^{2}$ defined by
\begin{align} % metric for resolved sphere
\de s_{\e}^{2}
=
\ell^{2} \lp \frac{1}{f_{\e} ( \vartheta )} \de \vartheta^{2} + q^{2} \sin^{2} \vartheta \de \tau^{2} \rp,
\end{align}
where a resolving function $f_{\e} ( \vartheta )$ satisfies
\begin{align}
f_{\e} ( \vartheta ) = \lc
	\begin{aligned}
	& \frac{1}{q^{2}} && \text{for } \vartheta \to 0, \pi, \\ %1
	& 1 && \text{for } \vartheta \in ( \e, \pi - \e ), %2
	\end{aligned} \right.
\end{align}
with a small $\e$. The zweibein $e^a$ and the spin connection $\om^{ab}$ on $S_{\e}^{2}$ are given by
\begin{align} % vielbein and spin connection
	\begin{aligned}
	e^{1} &= \frac{\ell}{\sqrt{f_{\e}}} \de \vartheta, \hspace{3em}
	e^{2} = \ell q \sin \vartheta \de \tau, \\ %1
	\om^{12} &= - \om^{21} = - q \sqrt{f_{\e}} \cos \vartheta \de \tau. %2
	\end{aligned}
\end{align}
The curvature $\Rcal_{\e}$ on $S_{\e}^{2}$ turns to be a nonsingular function due to the resolving function,
\begin{align} % curvature
\Rcal_{\e} = \frac{2}{\ell^{2}} \lp f_{\e} - \frac{1}{2} \cot \vartheta f_{\e}' \rp,
\end{align}
where $f_{\e}'$ represents the derivative with respects to $\vartheta$. Therefore, we consider the theory on $S_{q}^{2}$ as on $S_{\e}^{2}$ with taking the limit $\e \to 0$. Note that, for conformal field theories, the branched and the resolved two-sphere are thought of as the special cases of the squashed one. We can construct $\Ncal = ( 2, 2 )$ supersymmetry on $S_{\e}^{2}$ by making use of the results on the squashed two-sphere $S_{b}^{2}$ \cite{Gomis:2012wy}.

\paragraph{Killing spinors.} %%%%%%%%%%%%%%%%%%%%%%%%%%%%%%%%%%%%%%%%%%
Killing spinors, $\xi, \bxi$, on $S_{\e}^{2}$ can be obtained from the construction on general two-dimensional curved spaces discussed in \cite{Closset:2014pda, Bae:2015eoa}. They have studied $\Ncal =( 2, 2 )$ supersymmetric gauge theories with a vector-like R-symmetry and introducing the supergravity background (or the topological counterpart). In this setup, the generalized Killing spinor equations are acquired by the variations of gravitinos as
\begin{align} % general KSEs with a background field
	\begin{aligned}
	\lp \nabla_{\mu} + i V_{\mu} \rp \xi &= - \frac{1}{2} H \g_{\mu} \xi - \frac{i}{2} G \g_{\mu} \g^{3} \xi, \\ %1
	\lp \nabla_{\mu} - i V_{\mu} \rp \bxi &= - \frac{1}{2} H \g_{\mu} \bxi + \frac{i}{2} G \g_{\mu} \g^{3} \bxi, %2
	\end{aligned}
	\label{gkse}
\end{align}
where $\nabla_{\mu}$ is the standard spin connection, $V$ is a background gauge field for the R-symmetry, and $H, G$ are supergravity background fields. Now, we choose
\begin{align}
H = - \frac{i \sqrt{f_{\e}}}{\ell}, \hspace{2em}
G = 0.
\end{align}
Another choice also reproduces the corresponding Killing spinor equations on $S_{\e}^{2}$. Note that the supergravity background fields for unitary theories should be real in Lorentzian backgrounds. In Euclidean signature, however, they would be complex generally to keep supersymmetry. Therefore, the Killing spinor equations on $S_{\e}^{2}$ can be written by
\begin{align} % KSEs with a background field
	\begin{aligned}
	\Dcal_{\mu} \xi &= \lp \nabla_{\mu} + i V_{\mu} \rp \xi = + \frac{i \sqrt{f_{\e}}}{2 \ell} \g_{\mu} \xi, \\ %1
	\Dcal_{\mu} \bxi &= \lp \nabla_{\mu} - i V_{\mu} \rp \bxi = + \frac{i \sqrt{f_{\e}}}{2 \ell} \g_{\mu} \bxi. %2
	\end{aligned}
	\label{ksee}
\end{align}
Then, we set $V$ to be
\begin{align} % background gauge field
V = \frac{1 - q \sqrt{f_{\e}}}{2} \de \tau,
\label{bgss}
\end{align}
and consequently, the equations \eqref{ksee} have the solutions
\begin{align} % Killing spinors
\xi = e^{- i \frac{\tau}{2}} \begin{pmatrix} \sin \frac{\theta}{2} \\ - i \cos \frac{\theta}{2} \end{pmatrix}, \hspace{2em}
\bxi = e^{i \frac{\tau}{2}} \begin{pmatrix} \cos \frac{\theta}{2} \\ i \sin \frac{\theta}{2} \end{pmatrix},
\label{ksb2}
\end{align}
which are nothing but the Killing spinors on the round two-sphere $S^{2}$ \cite{Benini:2012ui} normalized as $\bxi \xi = - 1$. In other words, we can utilize \eqref{ksb2} even on $S_{\e}^{2}$ by choosing the specific value of $V$ \eqref{bgss}. In this paper, we treat the Killing spinors as Grassmann-odd variables. Note that the covariant derivative acts on the field components generically as
\begin{align}
\Dcal_{\mu} = \p_{\mu} + \frac{1}{4} \om_{\mu}^{a b} \g^{a b} - i \tilde{Q} A_{\mu} - i \tilde{R} V_{\mu},
\end{align}
where $\tilde{Q}$ is a charge for the gauge symmetry, and $\tilde{R}$ is a R-charge (the gauge field parts are replaced with its commutators when we consider a field in an adjoint representation). One can verify that $V$ is compatible with the integrability condition
\begin{align} % integrability condition
[ \Dcal_{\mu}, \Dcal_{\nu} ] \xi
= \lp \frac{1}{4} R_{\mu \nu a b} \g^{a b} + i \Vcal_{\mu \nu} \rp \xi,
\end{align}
where $\Vcal$ is the field strength of $V$.

\paragraph{Supersymmetry variations.} %%%%%%%%%%%%%%%%%%%%%%%%%%%%%%%%%%%%%
The vector multiplet consists of a gauge field $A$, two real scalars $\sg, \rho$, 2-component Dirac spinors $\la, \bla$, and an auxiliary scalar $D$. On the other hand, the chiral multiplet contains two complex scalars $\phi, \bphi$, 2-components Dirac spinors $\psi, \bpsi$, and auxiliary scalars $F, \bF$. $\Ncal = ( 2, 2 )$ supersymmetry for the vector multiplet on $S_{\e}^2$ can be constructed as
\begin{align} % SUSY for vector
	\begin{aligned}
	\dl A_{\mu} &= - \frac{i}{2} \lp \bxi \g_{\mu} \la - \bla \g_{\mu} \xi \rp, \\ %1
	\dl \rho &= - \frac{i}{2} \lp \bxi \g_{3} \la - \bla \g_{3} \xi \rp, \\ %2
	\dl \sg &= \frac{1}{2} \lp \bxi \la - \bla \xi \rp, \\ %3
	\dl \la &= i \g_{3} \xi \lp \frac{1}{2} \ve^{\mu \nu} F_{\mu \nu} + i [ \sg, \rho ] - \frac{\sqrt{f_{\e}}}{\ell} \rho \rp
	- \xi D
	%+ i \g^{\mu} \xi \lp \ve_{\mu \nu} \Dcal^{\nu} \eta + \Dcal_{\mu} \sg \rp, \\
	+ i \g^{\mu} \xi \Dcal_{\mu} \sg
	- \g_{3} \g^{\mu} \xi \Dcal_{\mu} \rho, \\ %4
	\dl \bla &= i \g_{3} \bxi \lp \frac{1}{2} \ve^{\mu \nu} F_{\mu \nu} - i [ \sg, \rho ] - \frac{\sqrt{f_{\e}}}{\ell} \rho \rp
	+ \bxi D
	%+ i \g^{\mu} \xi \lp \ve_{\mu \nu} \Dcal^{\nu} \eta - \Dcal_{\mu} \sg \rp, \\
	- i \g^{\mu} \bxi \Dcal_{\mu} \sg
	- \g_{3} \g^{\mu} \bxi \Dcal_{\mu} \rho, \\ %5
	\dl D &= - \frac{i}{2} \bxi \g^{\mu} \Dcal_{\mu} \la + \frac{i}{2} [ \bxi \la, \sg ] + \frac{1}{2} [ \bxi \g_{3} \la, \rho ]
	- \frac{i}{2} \Dcal_{\mu} \bla \g^{\mu} \xi + \frac{i}{2} [ \bla \xi, \sg ] + \frac{1}{2} [ \bla \g_{3} \xi, \rho ], %6
	\end{aligned} \label{susyvec}
\end{align}
where $F_{1 2} = \frac{1}{2} \ve^{\mu \nu} F_{\mu \nu}$ is the field strength of $A$, and for the chiral multiplet,
\begin{table}[t] % charge assignments
\begin{center}
\caption{The scaling weights and R-charges for the field contents.\label{dc}}
\vspace{-.5em}
	\begingroup
	\renewcommand{\arraystretch}{1.2}
	\begin{tabular}{|c||c|c||c|c|c|c|c|c||c|c|c|c|c|c|} \hline
	& $\xi$& $\bxi$ & $A_{\mu}$ & $\sg$ & $\eta$ & $\la$ & $\bla$ & $D$ & $\phi$ & $\bphi$ & $\psi$ & $\bpsi$ & $F$ & $\bF$ \\ \hline \hline
	scale & $- \frac{1}{2}$ & $- \frac{1}{2}$ & 1 & 1 & 1 & $\frac{3}{2}$ & $\frac{3}{2}$ & 2 & $\frac{\Dl}{2}$ & $\frac{\Dl}{2}$ & $\frac{\Dl + 1}{2}$ & $\frac{\Dl + 1}{2}$ & $\frac{\Dl + 2}{2}$ & $\frac{\Dl + 2}{2}$ \\ \hline % scale weight
	$\tilde{R}$ & $- 1$ & $1$ & 0 & 0 & $0$ & $- 1$ & 1 & 0 & $\Dl$ & $- \Dl$ & $\Dl - 1$ & $1 - \Dl$ & $\Dl - 2$ & $2 - \Dl$ \\ \hline % U(1)_R
	\end{tabular}
	\endgroup
\end{center}
\vspace{-1em}
\end{table}
\begin{align} % SUSY for chiral
	\begin{aligned}
	\dl \phi &= \bxi \psi, \\ %1
	\dl \bphi &= \bpsi \xi, \\ %2
	\dl \psi &= i \g^{\mu} \xi \Dcal_{\mu} \phi + i \xi \sg \phi + \g_{3} \xi \rho \phi - \frac{\Dl}{2 \ell} \sqrt{f_{\e}} \xi \phi
	+ \bxi F, \\ %3
	\dl \bpsi &= i \g^{\mu} \bxi \Dcal_{\mu} \bphi + i \bxi \bphi \sg - \g_{3} \bxi \bphi \rho - \frac{\Dl}{2 \ell} \sqrt{f_{\e}} \bxi \bphi
	+ \xi \bF, \\ %4
	\dl F &= \xi \lp i \g^{\mu} \Dcal_{\mu} \psi - i \sg \psi + \g_{3} \rho \psi - i \la \phi + \frac{\Dl}{2 \ell} \sqrt{f_{\e}} \psi \rp, \\ %5
	\dl \bF &= \bxi \lp i \g^{\mu} \Dcal_{\mu} \bpsi - i \bpsi \sg - \g_{3} \bpsi \rho + i \bphi \bla + \frac{\Dl}{2 \ell} \sqrt{f_{\e}} \bpsi \rp, %6
	\end{aligned} \label{susymat}
\end{align}
where we assign the R-charge $\Dl$ to the lowest component of the chiral multiplet (see Table \ref{dc}). When we treat supersymmetry as $\dl = \dl_{\xi} + \dl_{\bxi}$ which explcitly represent the transformations with respects of the corresponding Killing spinors, it is found that the supersymmetry algebra is closed as
\begin{align}
[ \dl_{\xi}, \dl_{\bxi} ]
=
\dl_{v} + \dl_{\La} + \dl_{R_{V}},
\end{align}
where $\dl_{v}$, $\dl_{\La}$, and $\dl_{R_{V}}$ are the traslation, the gauge transformation, and R-rotation, respectively, whose precise actions are summarized in Appendix \ref{Salg}. In addition, the remaining commutators vanish,
\begin{align}
[ \dl_{\xi_{1}}, \dl_{\xi_{2}} ]
=
[ \dl_{\bxi_{1}}, \dl_{\bxi_{2}} ]
=
0,
\end{align}
except for
\begin{align}
	\begin{aligned}
	\lb \dl_{\xi_{1}}, \dl_{\xi_{2}} \rb F
	&=
	- \Dl \xi_{[ 2} \g^{\mu} \g^{\nu} \Dcal_{\mu} \Dcal_{\nu} \xi_{1 ]} \phi
	+ i \Dl \xi_{[ 2} \g^{\mu \nu} \xi_{1 ]} \Vcal_{\mu \nu} \phi, \\ %1
	[ \dl_{\bxi_{1}}, \dl_{\bxi_{2}} ] \bF
	&=
	- \Dl \bxi_{[ 2} \g^{\mu} \g^{\nu} \Dcal_{\mu} \Dcal_{\nu} \bxi_{1 ]} \bphi
	- i \Dl \bxi_{[ 2} \g^{\mu \nu} \bxi_{1 ]} \Vcal_{\mu \nu} \bphi. %2
	\end{aligned}
\end{align}
These commutators vanish if the Killing spinors satisfy
\begin{align}
	\begin{aligned}
	\g^{\mu} \g^{\nu} \Dcal_{\mu} \Dcal_{\nu} \xi &= - \frac{1}{2} \lp \Rcal - 2 i \g^{\mu \nu} \Vcal_{\mu \nu} \rp \xi, \\ %1
	\g^{\mu} \g^{\nu} \Dcal_{\mu} \Dcal_{\nu} \bxi &= - \frac{1}{2} \lp \Rcal + 2 i \g^{\mu \nu} \Vcal_{\mu \nu} \rp \bxi. %2
	\end{aligned}
\end{align}
These sufficient conditions provide constraints to $f_{\e}$ as
\begin{align}
	\begin{aligned}
	\frac{1}{\ell^{2}} \lp \frac{i}{2} f'_{\e} \g^{1} - f_{\e} \rp \xi &= - \frac{1}{2} \lp \Rcal - 2 i \g^{\mu \nu} \Vcal_{\mu \nu} \rp \xi, \\ %1
	\frac{1}{\ell^{2}} \lp \frac{i}{2} f'_{\e} \g^{1} - f_{\e} \rp \bxi &= - \frac{1}{2} \lp \Rcal + 2 i \g^{\mu \nu} \Vcal_{\mu \nu} \rp \bxi. %2
	\end{aligned} \label{fadd}
\end{align}

With our supersymmetry \eqref{susyvec}, we can show that the Lagrangian $\Lcal_{\text{SYM}}$ for the vector multiplet is SUSY-exact, and the Fayet-Iliopoulos (FI) term $\Lcal_{\text{FI}}$ and the theta term $\Lcal_{\text{top}}$ are SUSY-invariant but not exact:
\begin{itemize} % each Lagrangian
\item The super-Yang-Mills (SYM) Lagrangian % SYM
	\begin{align}
	\Lcal_{\text{SYM}}
	&= - \dlz \dlbz \Tr \lb \frac{1}{2} \bla \la - 2 D \sg + \frac{\sqrt{f_{\e}}}{\ell} \sg^{2} \rb \non \\ %1
	&= \frac{1}{2} \Tr \lb \lp \frac{1}{2} \ve^{\mu \nu} F_{\mu \nu} - \frac{\sqrt{f_{\e}}}{\ell} \rho \rp^{2}
	+ \Dcal^{\mu} \sg \Dcal_{\mu} \sg
	+ \Dcal^{\mu} \rho \Dcal_{\mu} \rho
	- [ \sg, \rho ]^{2}
	+ D^{2} \right. \non \\ %2
	&\hspace{13em} + i \lp \bla \g^{\mu} \Dcal_{\mu} \la \rp
	+ i \bla [ \sg, \la ]
	+ \bla \g_{3} [ \rho, \la ] \Big]. %3
	\end{align}

\item The FI term with a FI parameter $\zeta$ % FI
	\begin{align}
	\Lcal_{\rm FI}
	=
	i \zeta \Tr \lb D - \frac{\sqrt{f_{\e}}}{\ell} \sg \rb.
	\end{align}
	
\item The theta term with a theta parameter $\theta$ % theta
	\begin{align}
	\Lcal_{\rm top}
	=
	- i \frac{\theta}{2 \pi} \Tr \lb F_{12} \rb.
	\end{align}

\end{itemize}
There also are the SUSY-exact Lagrangian $\Lcal_{\text{ch}}$ for the chiral multiplet and its mass term:
\begin{itemize} %each Lagrangian
\item The matter Lagrangian % matter
	\begin{align}
	\Lcal_{\text{ch}}
	&= - \dlz \dlbz \lb \bpsi \psi - 2 i \bphi \sg \phi + \frac{( \Dl - 1) \sqrt{f_{\e}}}{\ell} \bphi \phi \rb \non \\ %1
	&= \Dcal^{\mu} \bphi \Dcal_{\mu} \phi
	+ \bphi \sg^{2} \phi
	+ \bphi \rho^{2} \phi
	+ i \bphi D \phi
	+ \bF F
	+ i \frac{( \Dl - 1) \sqrt{f_{\e}}}{\ell} \bphi \sg \phi
	+ \frac{\Dl ( 2 - \Dl ) f_{\e}}{4 \ell^{2}} \bphi \phi \non \\ %2
	&\hspace{1em} - i \bpsi \g^{\mu} \Dcal_{\mu} \psi
	+ i \bpsi \sg \psi
	- \bpsi \g_{3}  \rho \psi
	+ i \bpsi \la \phi
	- i \bphi \bla \psi
	- \frac{\Dl \sqrt{f_{\e}}}{2 \ell} \bpsi \psi. %3
	\label{matLag}
	\end{align}

\item The mass term for the chiral multiplet % mass of the matter
	\begin{align}
	\Lcal_{\text{mass}}
	= \lb \bphi m^{2} \phi
	+ \bphi \lp 2 \sg + i \frac{( \Dl - 1 ) \sqrt{f_{\e}}}{\ell} \rp m \phi
	+ i \bpsi m \psi \rb.
	\label{matLag1}
	\end{align}
We can introduce the twisted mass $m$ for the chiral multiplet associated with the flavor symmetry $G_{\rm f}$ in the supersymmetric way \cite{Benini:2012ui, Doroud:2012xw}. As in other dimensions, we can accomplish it by weakly gauging $G_{\rm f}$ with the background gauge field, coupling the chiral multiplet to that gauge field, and tuning on the expectation value for the background fields. This is simply shifting the expectation value of the scalar field $\sg$ by the constant one taken in the Cartan subalgebra of $G_{\rm f}$. Namely, we turn on the twisted mass by
\begin{align}
\sg \to \sg + m,
\label{massshift}
\end{align}
where we omit the index for the flavor symmetry. Substituting this into \eqref{matLag} leads to the above mass term. As a result, the supersymmetry algebra enlarges to include this flavor symmetry. Since the background vector multiplet obeys the same supersymmetry transformations as \eqref{susyvec}, the mass term \eqref{matLag1} becomes $\dl$-exact by construction.

\end{itemize}

Let us comment on the imaginary constant shift of $\sigma$. This just corresponds to the R-charge, and if we turn off $\Dl$ in supersymmetry \eqref{susymat} and instead take
\begin{align}
\sg \to \sg + i \frac{\Dl}{2 \ell},
\end{align}
then we obtain the same Lagrangian as the original one \eqref{matLag}. Note that the flavor symmetry $G_{\rm f}$ is determined by the representation $\mathbf{R}$ of the gauge group under which the chiral multiplet transforms and the choice of the superpotential because it breaks the enhanced symmetry down to $G_{\rm f}$. For example, when $\mathbf{R}$ contains $N_{f}$ copies of an irreducible representation and there is the trivial superpotential, the theory has U$(N_{f})$ as part of the flavor symmetry. Accordingly, we have $N_{f}$ twisted masses $\overrightarrow{m} = ( m_{1}, m_{2}, \cdots, m_{N_{f}} )$ and $N_{f}$ U$(1)$ R-charges $\overrightarrow{\Dl} = ( \Dl_{1}, \Dl_{2}, \cdots, \Dl_{N_{f}} )$. To summarize, we can insert the holomorphic combination
\begin{align}
m_{I} + i \frac{\Dl_{I}}{2 \ell}
\end{align}
into the Lagrangian by shifting the expectation value of $\sg$. In this paper, however, we think of the R-charge $\Dl$ as an independent free parameter even in the presence of possible superpotentials.

%%%%%%%%%%%%%%%%%% section 2.2 %%%%%%%%%%%%%%%%%%%%%%%%%%%%%%%%
\subsection{Partition function}

\paragraph{Locus.} %%%%%%%%%%%%%%%%%%%%%%%%%%%%%%%%%%%%%%%%%%%%%%
The localized configuration for the field components can be derived from the positive definiteness of the SUSY-exact Lagrangians. The saddle point equations for the vector multiplet are found from $\Lcal_{\text{SYM}}$ as
\begin{align} % saddle point eq
0 = F_{12} - \frac{\sqrt{f_{\e}}}{\ell} \rho = \Dcal_{\mu} \sg = \Dcal_{\mu} \rho = [ \sg, \rho ] = D. \label{locus1}
\end{align}
The solution of these equations can be generically obtained as
\begin{align}
A = A_{\text{mon}}, \hspace{2em}
\rho = \frac{\text{\boldmath $s$}}{\ell q}, \hspace{2em}
\sg = - \text{\boldmath $a$}, \hspace{2em}
D = \la = \bla = 0,
\label{locusvec2}
\end{align}
where \text{\boldmath $a$} is a constant diagonal matrix, and $A_{\text{mon}}$ is the GNO monopole configuration defined by
\begin{align}
A_{\text{mon}} = \text{\boldmath $s$} \lp \kappa - \cos \vartheta \rp \de \tau, \hspace{2em}
\kappa = \lc
    \begin{aligned}
    + 1 & && \mbox{for } \vartheta \in [ 0, \pi / 2 ], \\
    - 1 & && \mbox{for } \vartheta \in [ \pi / 2, \pi ].
    \end{aligned} \right.
\end{align}
The matrix \text{\boldmath $s$} for the magnetic charge is taken in the Cartan subgroup of the gauge group with all half-integer components $\{ s_{i} \} \in \Zbb/2$.

In contrast, the locus of the chiral multiplet read from $\Lcal_{\text{ch}}$ is trivial, that is, all field contents vanish:
\begin{align}
\phi = \bphi = \psi = \bpsi = F = \bF = 0.
\label{locusmat2}
\end{align}

\paragraph{One-loop determinants.} %%%%%%%%%%%%%%%%%%%%%%%%%%%%%%%%%%%%%%
In computing the one-loop determinants, the boson and fermion eigenmodes which make a pair under a certain map give the trivial contribution because they are completely cancelled. The point of the calculation is to find the eigenmodes annihilated by such map but still satisfying the eigenvalue equations. To do that, we will accept the method that we directly solve the differential equations for the unpaired eigenmodes obtained by the eigenvalue equations as worked in \cite{Hama:2011ea, Gomis:2012wy, Tanaka:2013dca}.

Here, we only line up the final results of the one-loop determinants. The details for derivation are briefed in Appendix \ref{Vector} and \ref{Chiral}. The one-loop determinant $\Zcal_{\text{1-loop}}^{\text{vec}}$ for the vector multiplet by combining \eqref{vecfeigenv2} and \eqref{vecbeigenv2} is given by
\begin{align} % the one-loop determinant for the vector multiplet
\Zcal_{\text{1-loop}}^{\text{vec}}
&=
\frac{\det \Dl_{\rm vec}^{f}}{\det \dl_{\rm vec}^{b}} \non \\ %1
&\simeq
\prod_{\substack{ \alpha > 0 \\ \alpha ( s ) \neq 0}}
\lb \lp \alpha ( a ) \ell q \rp^{2} + \alpha ( s )^{2} \rb, %2
\label{vec1-loop1}
\end{align}
where $\simeq$ represents the equality up to the phase, and $\alpha$ is the root. Similarly, bringing \eqref{feigenv2} and \eqref{beigenv2} together results in the one-loop determinant $\Zcal_{\text{1-loop}}^{\text{ch}}$ for the chiral multiplet as
\begin{align} %\label{ch1-loop1} % the one-loop determinant for the chiral multiplet
\Zcal_{\text{1-loop}}^{\text{ch}}
&=
\frac{\det \Dl_{\text{ch}}^{f}}{\det \Dl_{\text{ch}}^{b}} \non \\ %1
&=
\prod_{w}
\prod_{j \geq 0}
\frac
{j + 1 + i w ( a ) \ell q + | w ( s ) | - \frac{\Dl}{2}}
{j - i w ( a ) \ell q + | w ( s ) | + \frac{\Dl}{2}}, %2
\end{align}
where $w$ is the wight vector. As in these works, to regulate the diverge product \eqref{ch1-loop1}, we use the Hurwitz zeta function which is one of the generalizations of the Riemann zeta function defined by \eqref{hurwitz}. Consequently, the one-loop determinant for the chiral multiplet is written by
\begin{align}
\Zcal_{\text{1-loop}}^{\text{ch}}
=
\prod_{w}
\frac
{\G \lp \frac{\Dl}{2} - i w ( a ) \ell q + | w ( s ) | \rp}
{\G \lp 1 - \frac{\Dl}{2} + i w ( a ) \ell q + | w ( s ) | \rp}.
\label{ch1-loop1}
\end{align}
Moreover, there is an additional phase factor for the contribution of the chiral multiplet derived by the index theorem \cite{Benini:2012ui, Doroud:2012xw}
\begin{align}
\lp - 1 \rp^{w ( s ) + | w ( s ) |}.
\end{align}
In fact, this can be absorbed into the Gamma functions in \eqref{ch1-loop1} so that the absolute value symbol can be removed, and the simplified one-loop determinant is
\begin{align}
\Zcal_{\text{1-loop}}^{\text{ch}}
=
\prod_{w}
\frac
{\G \lp \frac{\Dl}{2} - i w ( a ) \ell q - w ( s ) \rp}
{\G \lp 1 - \frac{\Dl}{2} + i w ( a ) \ell q - w ( s ) \rp}.
\label{ch1-loop2}
\end{align}
Surely, these contributions \eqref{vec1-loop1} and \eqref{ch1-loop2} are almost the same results as on the round sphere $S^{2}$ \cite{Benini:2012ui, Doroud:2012xw}, but they include an extra geometrical data $q$ on the $q$-covering space. This $q$-dependence can be considered as the specific effect from the conical singularities.
%We should notice that those contributions on $S_q^{2}$ are the same results as the ones on the round two-sphere $S^{2}$ \cite{Benini:2012ui, Doroud:2012xw} except for the factor $q$.

\paragraph{Exact partition function.} %%%%%%%%%%%%%%%%%%%%%%%%%%%%%%%%%%%%%%
We write down the remaining factors to finalize the partition function. The magnetic flux breaks the gauge symmetry $G$ down to the subgroup $H_{s}$, and the integral reduces to the one over the Cartan subalgebra $\mathfrak{t}$. %constrained by $[ a, s ]$.
This argument generates the Jacobian $\Jcal$ called the Vandermonde determinant in the integration measure \cite{Benini:2012ui, Doroud:2012xw},
\begin{align}
\Jcal ( a, s )
= \frac{1}{| W ( H_{s} ) |} \prod_{\substack{\alpha > 0 \\ \alpha ( s ) = 0}} \alpha ( a )^{2},
\label{vande}
\end{align}
where $W ( H_{s} )$ is the Weyl group of $H_{s}$.

The FI term and the theta term also contribute to the partition function as classical ones evaluated only by substituting the locus \eqref{locusvec2},
\begin{align}
S_{\rm FI}
&=
i \zeta \int {\rm d}^{2} x \sqrt{g}\ \Tr \lb D - \frac{\sqrt{f_{\e}}}{\ell} \sg \rb
=
4 i \pi \zeta \ell q \Tr \lb \text{\boldmath $a$} \rb, \\ %1
S_{\rm top}
&=
- i \frac{\theta}{2 \pi} \int {\rm d}^{2} x \sqrt{g}\ \Tr \lb F_{12} \rb
=
- 2 i \theta \Tr \lb \text{\boldmath $s$} \rb. %2
\end{align}
Accordingly, the partition function $\Zcal_{q}$ on the branched two-sphere $S_{q}^2$ combining all factors is obtained as
%\begin{align} \label{exactPF1} % exact PF 1
%\Zcal_{q} ( \zeta, \theta, \Dl )
%&=
%\sum_{\{ s_i \} \in \Zbb/2}
%\frac{1}{| W ( H_{s} ) |}
%\int_{\mathfrak{t}} [ \de a ]
%e^{- 4 i \pi \zeta \ell q \Tr \lb \text{\boldmath $a$} \rb + 2 i \theta \Tr \lb \text{\boldmath $s$} \rb}
%\prod_{\alpha > 0}
%\lb \lp \alpha ( a ) \ell q \rp^{2} + \alpha ( s )^{2} \rb \non \\ %1
%&\hspace{7em} \times
%\prod_{I, w_{I}} \lp - 1 \rp^{w_{I} ( s ) + | w_{I} ( s ) |}
%\frac
%{\G \lp \frac{\Dl}{2} - i w_{I} ( a ) \ell q + | w_{I} ( s ) | \rp}
%{\G \lp 1 - \frac{\Dl}{2} + i w_{I} ( a ) \ell q + | w_{I} ( s ) | \rp}, %2
%\end{align}
%where the index $I$ runs for the number of flavors. In fact, the phase factor in the second line of \eqref{exactPF1} can be absorbed into the Gamma functions so that the absolute value symbol can be removed. Thus, the simplified version of the partition function is
\begin{align} \label{exactPF2} % exact PF 2
\Zcal_{q} ( \zeta, \theta, \Dl )
&=
\sum_{\{ s_i \} \in \Zbb/2}
\frac{1}{| W ( H_{s} ) |}
\int_{\mathfrak{t}} [ \de a ]
e^{- 4 i \pi \zeta \ell q \Tr \lb \text{\boldmath $a$} \rb + 2 i \theta \Tr \lb \text{\boldmath $s$} \rb}
\prod_{\alpha > 0}
\lb \lp \alpha ( a ) \ell q \rp^{2} + \alpha ( s )^{2} \rb \non \\ %1
&\hspace{12.5em} \times
\prod_{I, w_{I}}
\frac
{\G \lp \frac{\Dl}{2} - i w_{I} ( a ) \ell q - w_{I} ( s ) \rp}
{\G \lp 1 - \frac{\Dl}{2} + i w_{I} ( a ) \ell q - w_{I} ( s ) \rp}, %2
\end{align}
where the index $I$ runs for the number of flavors. The expression \eqref{exactPF2} includes the branching parameter $q$ since the one-loop determinants themselves depend on it, but actually we can remove it (up to the overall constant) by rescaling the Coulomb moduli such that $\mathbf{a} \to \mathbf{a}/q$. We notice that if the twisted mass is added as explained in \eqref{massshift}, $q$ multiplied by the mass remains in the partition function. This point become much clear when we move to the formula by Higgs branch localization shown below.

\paragraph{Vortex partition function.} %%%%%%%%%%%%%%%%%%%%%%%%%%%%%%%%%%%%%%
The partition function on $S^2$ computed by Coulomb branch localization is found to be equivalent to the vortex partition function \cite{Dimofte:2010tz} which results from Higgs branch localization \cite{Benini:2012ui, Doroud:2012xw}. To see this observation on $S_{q}^2$, let us consider an U$(1)$ gauge theory with $N_{f}$ fundamental and $N_{a}$ anti-fundamental matters. The partition function of this theory is given by
\begin{align} \label{nfnaPF1} % PF
\Zcal_{q}^{( N_{f}, N_{a} )} ( \zeta, \theta; m, \tilde{m} )
=
\sum_{s \in \Zbb/2}
e^{2 i \theta s}
\int \frac{\de a}{2 \pi} e^{- 4 i \pi \zeta a q}
\prod_{i = 1}^{N_{f}}
\frac{\G ( - i a q - i m_{i} q - s )}{\G ( 1 + i a q + i m_{i} q - s )}
\prod_{\bar{\imath} = 1}^{N_{a}} 
\frac{\G ( i a q - i \tilde{m}_{\bar{\imath}} q + s )}{\G ( 1 - i a q + i \tilde{m}_{\bar{\imath}} q + s )},
\end{align}
where we set $( \Dl, \ell ) = ( 0, 1 )$ for simplicity, and $m_{i}$ ($\tilde{m}_{\bar{\imath}}$) is a twisted mass of each fundamental (anti-fundamental). Now, suppose $N_{f} > N_{a}$, or $N_{f} = N_{a}$ and $\zeta > 0$. The $l$-th tower of poles coming from the numerator of the contributions for the fundamentals are
\begin{align} \label{nfnapoles} % poles
a_{l, k} q = - m_{l} q - i k - i | s |,
\end{align}
with $k \in \Zbb_{\geq 0}$. %It is useful to define the following combinations of twisted masses:
%\begin{align} % Def of symbols
%M_{i}^{l} &= m_{i} - m_{l}, \hspace{2.5em}
%\widetilde{M}_{\bar{\imath}}^{l} = \tilde{m}_{\bar{\imath}} + m_{l}.
%\end{align}
Evaluating the residues at the poles \eqref{nfnapoles} with setting $M_{i}^{l} = m_{i} - m_{l}$ and $\widetilde{M}_{\bar{\imath}}^{l} = \tilde{m}_{\bar{\imath}} + m_{l}$ results in the partition function in the factorized form,
\begin{align}
\Zcal_{q}^{( N_{f}, N_{a} )} ( \zeta, \theta; m, \tilde{m} )
&=
\sum_{l = 1}^{N_{f}}
e^{4 i \pi \zeta q m_{l}}
Z_{\text{1-loop}}^{(l)} ( q M_{i}^{l}, q \widetilde{M}_{\bar{\imath}}^{l} ) \notag \\ %1
%Z_{\text{v}}^{(l)} ( z )
%Z_{\text{av}}^{(l)} ( \bar{z} ),
&\hspace{2.5em} \times
Z_{\rm vortex}^{{\rm U}(1)} ( - z, 1, - i q M_{i ( \neq l )}^{l}, - i q \widetilde{M}_{\bar{\imath}}^{l} )
Z_{\rm vortex}^{{\rm U}(1)} ( - \bar{z}, - 1, i q M_{i ( \neq l )}^{l}, i q \widetilde{M}_{\bar{\imath}}^{l} ), %2
\end{align}
where $z := e^{- 2 \pi \zeta + i \theta}$, and the last three factors are given by
\begin{align} % vortex partition function
Z_{\text{1-loop}}^{(l)} ( q M_{i}^{l}, q \widetilde{M}_{\bar{\imath}}^{l} )
&=
\prod_{\substack{i = 1 \\ i \neq l}}^{N_{f}}
\frac{\G ( - i M_{i}^{l} )}{\G ( 1 + i M_{i}^{l} )}
\prod_{\bar{\imath} = 1}^{N_{a}}
\frac{\G ( - i \widetilde{M}_{\bar{\imath}}^{l} )}{\G ( 1 + i \widetilde{M}_{\bar{\imath}}^{l} )}, \\ %1 
Z_{\rm vortex}^{{\rm U}(1)} ( z, \ve, m_{i}, \tilde{m}_{\bar{\imath}} )
&=
\sum_{k \geq 0} \frac{z^{k}}{\ve^{( N_{f} - N_{a} ) k} k!}
\frac{\prod_{\bar{\imath} = 1}^{N_{a}} \lp \frac{\tilde{m}_{\bar{\imath}}}{\ve} \rp_{k}}
{\prod_{i = 1}^{N_{f} - 1} \lp \frac{m_{i}}{\ve} - k \rp_{k}}, \label{vortexpfu1} %2
\end{align}
with the shifted factorial $( a )_{k} := \prod_{n = 0}^{k - 1} ( a + n)$. The contribution \eqref{vortexpfu1} is nothing but the vortex partition function for an U$(1)$ gauge group with $( N_{f}, N_{a} )$ flavors in $\Om$-background \cite{Dimofte:2010tz, Benini:2012ui}. 
%where $z := e^{- 2 \pi \zeta + i \theta}$. The last two contributions are nothing but the ones given in \cite{Benini:2012ui} which can be naturally identified with the vortex partition function for an U$(1)$ gauge group with $( N_{f}, N_{a} )$ flavors in $\Om$-background \cite{Dimofte:2010tz} (see \cite{Benini:2012ui} for precise expressions).
%Hence, all results from the calculation above are absolutely compatible with literature, and what we should emphasize here is that the partition function on the branched sphere $S_{q}^2$ does really \textit{not} depend on the parameter $q$.
All results from the calculation above are basically compatible with literature, and the extra parameter $q$ appears as an effect of the conical singularities at the north and the south pole if we switch on the twisted masses\footnote{The author would like to thank Bruno Le Floch for indicating a typo here in the first version.}.

\subsection{Supersymmetric R\'enyi entropy}
We use the exact partition function $\Zcal_{q}$ on the branched sphere to obtain the supersymmetric R\'enyi entropy $\Scal_{q}^{\text{susy}}$ \eqref{sre} in two dimensions as mentioned in Introduction. For our result \eqref{exactPF2}, the branching parameter $q$ appears always in the way to attach to the Coulomb moduli $\mathbf{a}$, but the redefinition of $\mathbf{a}$ as $\mathbf{a} \to \mathbf{a}/q$ can remove $q$-dependence from the partition function (up to the overall constant from the integration measure), and also $\Zcal_{q}$ of the system without the vector multiplet does not depend on $q$ since $q$ is attached only with the Coulomb moduli. On the other hand, as described above, introducing the twisted mass brings $q$-dependence into the partition function. We should stress that our results are physically secure only in CFTs, which means that the scaling dimension of the field is allowed to be a certain value. Because the presence of the twisted mass in general breaks conformal invariance, we do not take it into account in discussing the physical quantity. Therefore, $\Zcal_{q}$ does not essentially depend on $q$ and is equal to $\Zcal_{q = 1}$ on the round sphere. Actually, this might be expected because it is shown that physical quantities respecting the conformal symmetry do not depend on the deformation parameter of the two-sphere \cite{Gomis:2012wy}. That observation means that the supersymmetric R\'enyi entropy simply reduces to the ordinary entanglement entropy in two dimensions. The result we got is just consistent with the one of main results in \cite{Giveon:2015cgs} that the SRE with a single interval on the spatial direction is independent of the number of sheets used in the replica trick and equivalent to the standard entanglement entropy. This is because the SRE on the branched sphere can be regarded as the one mapped conformally from one-dimensional space with one interval.
\section{Defect interpretation} \label{Defect}
In this section, we would like to give the defect operator interpretation to our calculation on the branched sphere $S_{q}^{2}$ as explained in \cite{Nishioka:2013haa}. In this picture, the defects are located on the entangling surface, namely, they are codimension-2 and set singular boundary conditions on the fields near them. We can see the nontrivial effect of the defects in free field theories \cite{Casini:2009sr} as follows. We split a filed $\Phi$ defined on the $q$-covering space into $\{ \Phi_{n} \}_{n = 1}^{q}$ on each sheet. The field $\Phi_{n}$ on the right side of each sheet should be simply connected with $\Phi_{n + 1}$ on the left side of the next sheet. These boundary conditions are written in an unified form as a matrix $T$. In fact, the matrix $T$ can be diagonarized by fields $\{ \widetilde{\Phi}_{n} \}_{n = 1}^{q}$ defined from $\{ \Phi_{n} \}_{n = 1}^{q}$ twisted by monodromy around the defects placed on the entangling surface. Therefore, the one-loop determinant for $\Phi$ may be recast as the contributions of $q$ fields $\widetilde{\Phi}_{n}$ in the defect background.

To confirm this interpretation of our results, we start with briefly summarizing the sphere partition functions in the presence of the defects conjecturally derived from the orbifolding action on the squashed sphere \cite{Hosomichi:2015pia}. Then, we show interpolating the effect of the defects and the conical singularities in terms of the one-loop determinants.

%%%%%%%%%%%%%%%%%% section 3.1 %%%%%%%%%%%%%%%%%%%%%%%%%%%%%%%%
\subsection{Partition function with defects}
In this paper, we concentrate only on the defects for a dynamical gauge field $A$, which located on the north (N) and the south (S) pole of a two-sphere in order to keep supersymmetry (see \cite{Okuda:2015yra} for the argument of the defects for a background gauge field). Those are expressed as the non-vanishing profile of $A$ at the poles \cite{Hosomichi:2015pia},
\begin{align}
A \simeq \lc
	\begin{aligned}
	& \text{\boldmath $\eta$}^{\text{\tiny N}} \de \varphi && \text{at } \vt = 0, \\
	& \text{\boldmath $\eta$}^{\text{\tiny S}} \de \varphi && \text{at } \vt = \pi,
	\end{aligned} \right.
\RA
F_{1 2} \simeq \lc
	\begin{aligned}
	+ & 2 \pi \text{\boldmath $\eta$}^{\text{\tiny N}} \dl_{\text{\tiny N}}^{2}, \\
	- & 2 \pi \text{\boldmath $\eta$}^{\text{\tiny S}} \dl_{\text{\tiny S}}^{2},
	\end{aligned} \right.
\end{align}
where $\dl_{\text{\tiny N,S}}^{2} := \dl_{\text{\tiny N,S}}^{2} ( x_{i} )$ is a delta function on the pole with local Cartesian coordinates $( x^{1}, x^{2} )$. The matrices $\text{\boldmath $\eta$}^{\text{\tiny  N,S}}$ are holonomies (called vorticities) around the poles which are embedded  into the Cartan subgroup of the gauge group. This fact is concluded by the gauge invariance to connect the north and south patch of the sphere.

\paragraph{Twisted boundary conditions.} %%%%%%%%%%%%%%%%%%%%%%%%%%%%%%%%%%%
The defects we are considering are just local singularities corresponding to the excitation of local operators with infinite masses. Consequently, it is natural to think that the theory with the defects can be described as the one defined on a background with some singularities at the poles instead of the defects. Actually, the equivalence between the gauge theories containing the defects and the ones defined on manifolds divided by orbifolds has been proposed by \cite{biswas1997}. This idea roughly can be understood in terms of the boundary condition imposed on the fields as follows \cite{Hosomichi:2015pia} : to make the discussion comprehensive, we focus on an U$(1)$ gauge group and the special value
\begin{align}
\eta^{\text{\tiny N}}
=
\eta^{\text{\tiny S}}
=
\eta
=
\frac{r}{K},
\end{align}
where $r = 0, 1, \cdots, K - 1$, and $K \in \Zbb$. For these charges, we can remove the singularity of the gauge field at the poles by implementing the unusual gauge transformation
\begin{align}
A \to A' = A - \eta \de \varphi.
\label{ungauge}
\end{align}
This leads to breaking the single-valuedness of the charged matter, namely, the matter $\Phi ( \vt, \varphi )$ with gauge charge $+ 1$ obeys the twisted boundary condition
\begin{align}
\Phi ( \vt, \varphi + 2 \pi )
=
e^{- 2 \pi i \eta}
\Phi ( \vt, \varphi ).
\label{vortexbc1}
\end{align}
As a result, the $\eta$-dependence is naturally encoded into the one-loop determinant. On the other hand, in the gauge theories on the orbifolded sphere $S^{2}/\Zbb_{K}$, the $\Zbb_{K}$ symmetry acts on the charged matter $\Phi$ as a gauge rotation such that
\begin{align}
\Phi ( \vt, \varphi + \frac{2 \pi}{K} )
=
e^{- 2 \pi i \frac{r}{K}}
\Phi ( \vt, \varphi ).
\end{align}
This is a similar observation as with the defects \eqref{vortexbc1}, then it is reasonable that we expect that the defects on the poles can be recast as geometrical singularities on there.

From this point of view, the partition function on the squashed sphere $S_{b}^{2}$ in the presence of the defects could be obtained as the one on the orbifolded sphere $S^{2}/\Zbb_{K}$ by identifying the SUSY-preserving twisted boundary conditions for the fields (eq.(4.14) in \cite{Hosomichi:2015pia}). Indeed, such boundary conditions are sensitive to R-charges and spins of the fields as well as gauge charges since the Killing spinors are charged under the $\Zbb_k$ rotation. This twisting brings nontrivial effects into the one-loop determinant for the chiral multiplet.

\paragraph{Exact results with defects.} %%%%%%%%%%%%%%%%%%%%%%%%%%%%%%%%%%%%%
As explained in \cite{Hosomichi:2015pia}, we can deal with the general situation $\text{\boldmath $\eta$}^{\text{\tiny N}} \neq \text{\boldmath $\eta$}^{\text{\tiny S}}$ where the twisting effects of these vorticities are differently encoded into the boundary condition. For the vector multiplet, we take an U$(N)$ gauge symmetry as a concrete example, and the vorticities are expressed by
\begin{align}
\text{\boldmath $\eta$}^{\text{\tiny N}}
=
\diag ( \eta_{1}^{\text{\tiny N}}, \eta_{2}^{\text{\tiny N}}, \cdots, \eta_{N}^{\text{\tiny N}} ), \hspace{2em}
\text{\boldmath $\eta$}^{\text{\tiny S}}
=
\diag ( \eta_{1}^{\text{\tiny S}}, \eta_{2}^{\text{\tiny S}}, \cdots, \eta_{N}^{\text{\tiny S}} )
\end{align}
subjected to the flux quantization condition $2 \text{\boldmath $s$} - \text{\boldmath $\eta$}^{\text{\tiny N}} + \text{\boldmath $\eta$}^{\text{\tiny S}} \in \Zbb^{N}$. Then, the one-loop determinant for the vector multiplet is obtained as
\begin{align} \label{vec1-loopdefect} % vector with defects
Z_{\text{1-loop}}^{\text{vec}} ( \text{\boldmath $\eta$}^{\text{\tiny N}}, \text{\boldmath $\eta$}^{\text{\tiny S}} )
&=
( - 1 )^{\frac{1}{2} N ( N - 1 )}
( - 1 )^{( N - 1 ) \Tr ( 2 \text{\boldmath $s$} - \text{\boldmath $\eta$}^{\text{\tiny N}} + \text{\boldmath $\eta$}^{\text{\tiny S}} )} \notag \\ %1
&\hspace{1em} \times
\sg ( \text{\boldmath $\eta$}^{\text{\tiny N}} )
\sg ( \text{\boldmath $\eta$}^{\text{\tiny S}} )
\prod_{\substack{a < b \\[.1em] \eta_{ab}^{\text{\tiny N}} = 0}} ( s_{ab} - i \ell a_{ab} )
\prod_{\substack{a < b \\[.1em] \eta_{ab}^{\text{\tiny S}} = 0}} ( - s_{ab} - i \ell a_{ab} ), %2
\end{align}
where $X_{ab} := X_{a} - X_{b}$, and the parity factor $\sg ( \text{\boldmath $\eta$} )$ of the unique permutation $\pi ( a )$ acts on \text{\boldmath $\eta$} such that
\begin{align}
\pi ( a ) < \pi ( b)
\RA
\lp \eta_{\pi ( a )} < \eta_{\pi ( b )} \rp
\text{ or }
\lp \eta_{\pi ( a )} = \eta_{\pi ( b )} \text{ for } a < b \rp.
\end{align}
We would like to comment on two things about the vector multiplet sector. The first one is to note that the one-loop determinant $Z_{\text{1-loop}}^{\text{vec}} ( \text{\boldmath $\eta$}^{\text{\tiny N}}, \text{\boldmath $\eta$}^{\text{\tiny S}} )$ does not contain the dependence on the vorticities inside the products because we remove the singularities of the gauge field by the gauge transformation \eqref{ungauge}. The second one lies in the degeneration of eigenvalues of $\text{\boldmath $\eta$}^{\text{\tiny N,S}}$. As one can see, $Z_{\text{1-loop}}^{\text{vec}} ( \text{\boldmath $\eta$}^{\text{\tiny N}}, \text{\boldmath $\eta$}^{\text{\tiny S}} )$ shows nontrivial contributions if some of the eigenvalues of $\text{\boldmath $\eta$}^{\text{\tiny N,S}}$ degenerate, that is, $\eta_{ab}^{\text{\tiny N,S}} = 0$. This is in fact related to which Levi subgroup $\Lbb$ preserved by the defects we choose.

The localization calculation under the twisted boundary conditions provides the one-loop determinant for the chiral multiplet
\begin{align} \label{ch1-loopdefect} % chiral with defects
Z_{\text{1-loop}}^{\text{ch}} ( \text{\boldmath $\eta$}^{\text{\tiny N}}, \text{\boldmath $\eta$}^{\text{\tiny S}} )
&=
( - 1 )^{\sum_{w} [ w ( \eta^{\text{\tiny S}} ) ]_{\Delta}}
\prod_{w}
\frac
{\G ( \frac{\Delta}{2} + w ( s ) - w ( \eta^{\text{\tiny N}} ) - i w ( a ) \ell + [ w ( \eta^{\text{\tiny N}} ) ]_{\Delta} )}
{\G ( 1 - \frac{\Delta}{2} + w ( s ) + w ( \eta^{\text{\tiny S}} ) + i w ( a ) \ell - [ w ( \eta^{\text{\tiny S}} ) ]_{\Delta} )} \notag \\ %1
&=
( - 1 )^{\sum_{w} \lc 2 w ( s ) + w ( \eta^{\text{\tiny S}} ) - w ( \eta^{\text{\tiny N}} ) \rc }
( - 1 )^{[ w ( \eta^{\text{\tiny N}} ) ]_{\Delta}} \notag \\ %2
&\hspace{1em} \times
\prod_{w}
\frac
{\G ( \frac{\Delta}{2} - w ( s ) - w ( \eta^{\text{\tiny S}} ) - i w ( a ) \ell + [ w ( \eta^{\text{\tiny S}} ) ]_{\Delta} )}
{\G ( 1 - \frac{\Delta}{2} - w ( s ) + w ( \eta^{\text{\tiny N}} ) + i w ( a ) \ell - [ w ( \eta^{\text{\tiny N}} ) ]_{\Delta} )}, %3
\end{align}
where the symbol $[ \eta ]_{\Delta}$ represents an integer-valued function combining a ceiling function and a floor function as
\begin{align}
[ \eta ]_{\Delta}
= \lc
	\begin{aligned}
	& \lceil \eta - {\textstyle \frac{\Delta}{2}} \rceil && \text{for } \Delta < 1, \\ %1
	& \lfloor \eta - {\textstyle \frac{\Delta}{2}} + 1 \rfloor && \text{for } \Delta > 1. %2
	\end{aligned} \right.
	\label{cffn}
\end{align}
Note that the sign factor in \eqref{ch1-loopdefect} is selected in the way that the antipodal map under which the sign of $\text{\boldmath $s$}$ is flipped and $\text{\boldmath $\eta^{\text{\tiny N,S}}$}$ are exchanged is still an anomalous symmetry. In what follows, we accept the second line of \eqref{ch1-loopdefect} for our purpose to compare the contributions in the presence of the defects with the results on the branched sphere.

%%%%%%%%%%%%%%%%%% section 3.2 %%%%%%%%%%%%%%%%%%%%%%%%%%%%%%%%
\subsection{Interplay of defects and conical singularities}
Let us revisit the exact results on the branched sphere $S_{q}^2$ and indicate how the defect interpretation of them works.

\paragraph{Vector multiplets.} %%%%%%%%%%%%%%%%%%%%%%%%%%%%%%%%%%%%%%%%%
We can easily see that the one-loop determinant \eqref{vec1-loop1} for the vector multiplet in the U$(N)$ gauge group combined with the Vandermonde determiant \eqref{vande} is rewritten as
\begin{align} % the one-loop determinant for the vector multiplet
\Zcal_{\text{1-loop}}^{\text{vec}}
&=
\prod_{a < b}
\lb \lp \ell q a_{ab} \rp^{2} + s_{ab}^{2} \rb \notag \\ %1
&=
\prod_{a < b}
\lp i \ell q a_{ab} + s_{ab} \rp
\prod_{a < b}
\lp - i \ell q a_{ab} + s_{ab} \rp. %2
\label{vec1-loop2}
\end{align}
This is the vector contribution with the defects \eqref{vec1-loopdefect} up to the sign factor in the case where the vorticities at the north and the south pole of $S_{q}^2$ are the same to be proportional to the identity matrix $\mathbf{1}$,
\begin{align}
\text{\boldmath $\eta$}^{\text{\tiny N}}
=
\text{\boldmath $\eta$}^{\text{\tiny S}}
=
\eta \times \mathbf{1}.
\label{eta1}
\end{align}
The condition \eqref{eta1} is natural to be taken since intensity of the conical singularities at both poles is identical, and the scalar curvature is a value independent of the choice of the gauge group. The above expression contains only one vector multiplet, however, it is seemingly not compatible with the situation explained in \cite{Nishioka:2013haa} where the contribution for the vector on the $q$-covering space is given by gluing the contributions for $q$ vector multiplets on the defect background (eq.(4.18) in \cite{Nishioka:2013haa}).
%{\color{red} it is seemingly not compatible with the situation in the presence of the defects explained in \cite{Nishioka:2013haa} where the contribution for the vector on the $q$-covering space are glued by the contributions for $q$ vector multiplets on the defect background.}
Nevertheless, we can recast \eqref{vec1-loop2} as that for the vector with the existence of the defects. For the defects under consideration, we take away the singularities of the gauge field by the irregular gauge transformation \eqref{ungauge}. Moreover, the flux quantization condition is set on the $q$-covering space, which means that $2 s_{a}$ are still integers and consistent with the condition on $S^{2}/\Zbb_{K}$. Therefore, it still can be regarded as the contribution \eqref{vec1-loopdefect} for a single vector multiplet consistently defined on the defect background. Note that this discussion is special for two dimensions differently from higher dimensions in which the codimension-2 defects are non-local. For general non-local defects, we cannot naively take the higher form extension of the gauge transformation \eqref{ungauge}. In summary, the vector multiplet on $S_{q}^2$ can be interpreted as the one on $S^2$ with the defects having the vorticities \eqref{eta1} under the gauge transformation \eqref{ungauge}.

\paragraph{Chiral multiplets.} %%%%%%%%%%%%%%%%%%%%%%%%%%%%%%%%%%%%%%%%%
On the other hand, the one-loop determinant \eqref{ch1-loop2} for a single chiral multiplet with R-charge $\Dl$ can be re-expressed using the difference equation \eqref{pndifference} and the multiplication theorem \eqref{multiGL} of the Gamma function so that
\begin{align}
\Zcal_{\text{1-loop}}^{\text{ch}}
&=
\prod_{w}
\frac
{\G \lp \frac{\Dl}{2} - i w ( a ) \ell q - w ( s ) \rp}
{\G \lp 1 - \frac{\Dl}{2} + i w ( a ) \ell q - w ( s ) \rp} \notag \\ %1
&=
q^{\sum_{w} \lc \Dl - 2 i w ( a ) \ell q - 1 \rc}
\prod_{k = 0}^{q - 1}
\prod_{w}
\frac
{\G \lp \frac{\Dl}{2 q} - i w ( a ) \ell - \frac{w ( s )}{q} + \frac{k}{q} \rp}
{\G \lp 1 - \frac{\Dl}{2 q} + i w ( a ) \ell - \frac{w ( s )}{q} - \frac{k}{q} \rp}. %2
\label{gl1}
\end{align}
Actually, as comparing with the result in the presence of the defects \eqref{ch1-loopdefect}, the rhs of \eqref{gl1} is the collection of a contribution for a chiral multiplet on each sheet with a vorticity $\eta_{\text{ch}}$,
\begin{align}
	\begin{aligned}
	\text{\boldmath $\eta$}^{\text{\tiny N}}
        &=
        \text{\boldmath $\eta$}^{\text{\tiny S}}
        =
        \eta_{\text{ch}} \times \mathbf{1}, \\
	\eta_{\text{ch}}
        &=
        \frac{\Dl}{2} \lp 1 - \frac{1}{q} \rp - \frac{k}{q},
        \hspace{2em}
        k = 0, 1, \dots, q - 1.
	\end{aligned}
\label{vorticitych}
\end{align}
The $\Dl$-dependence should be encoded into $\eta_{\text{ch}}$ because of twisting the supercharges by introducing the background field $V$ \eqref{bgss}. Also, we can identify the factor $( 1 - \frac{1}{q} )$ with the coefficient of the curvature singularity \eqref{qcurvature}. Moreover, since we take the usual flux quantization on the $q$-covering space, each contribution depends on a fractional magnetic charge divided by $q$.

Although the conical singularities of $S_{q}^2$ may be translated into the language of the vorticity, there are two obstructions remaining to claim the correspondence of the chiral multiplets between in two pictures. One thing is that the function $[ \eta ]_{\Dl}$ \eqref{cffn} could not appear in the one-loop determinant \eqref{ch1-loop1} on $S_{q}^2$. This discrepancy is originated from the fact that the twisted boundary conditions are imposed on the fields on the defect background, but not on the branched sphere\footnote{The author is thankful to Kazuo Hosomichi who has pointed out this.}. The other is the existence of the prefactor in the second line of \eqref{gl1} which includes the linear dependence on $a$. This dependence actually changes the final value of the integral for the partition function. Note that the linear $\Dl$-dependence also appears in the prefactor, but because it is constant unlike the $a$-dependence, we neglect it for the present.

To resolve these points, we should notice that our results on the branched sphere $S_{q}^2$ computed by the localization may be available for the specific values of R-charges corresponding to the theories which flow to superconformal field theories in IR. We expect that the defect interpretation of our geometrical singularities works only in that case. To confirm this statement, we calculate the R-charges for some gauged linear sigma models (GLSMs) whose low energy theories describe Calabi-Yau (CY) manifolds as target spaces by utilizing $c$-maximization \cite{Benini:2012cz, Benini:2013cda} in Appendix \ref{Ex22}. It is found from simple computations that such R-charges satisfy the condition $0 < \Dl < 1$ in all cases, and, as a result, $[ \eta_{\text{ch}} ]_{\Dl} = 0$ for small $\eta_{\text{ch}}$. This somehow supports the thing commented in \cite{Hosomichi:2015pia} that the twisted boundary conditions do not give the effect of introducing the factor $[ \eta ]_{\Dl}$ in the one-loop determinant if the vorticity is small. Thus, the defect expression \eqref{gl1} of our result should not contain the integer-valued function $[ \eta ]_{\Dl}$ in the region of the R-charge where the localization calculation becomes reliable.
Also, since these theories must be non-anomalous for the gauge symmetry, the $a$-dependent part of the prefactor in \eqref{gl1} are completely cancelled out with combining all matter contents.
As a consequence, the contributions \eqref{vec1-loop1} and \eqref{ch1-loop2} for the field contents on the $S_{q}^{2}$ can be translated into the languages of the defect background \eqref{vec1-loopdefect} and \eqref{ch1-loopdefect} for the specific R-charge with which the theory exhibits superconformal symmetry. In conclusion, we can provide the description of the superconformal field theories defined on the branched sphere as in the presence of the defects with equal and small vorticities located on the north and the south pole.

%%%%%%%%%%%%%%%%%% section 4 %%%%%%%%%%%%%%%%%%%%%%%%%%%%%%%%%
\section{Discussions} \label{Discussions}
We derive the exact formulas on the $q$-branched two-sphere. The one-loop determinant for each multiplet itself has the dependence on the parameter $q$, whereas it is found that the partition function is essentially independent of $q$. Consequently, the supersymmetric R\'enyi entropy defined by it becomes equivalent to the usual entanglement entropy. We also give the defect interpretation to our results and show that it can work when we consider the theories with an appropriate R-charge which flow to superconformal theories. However, there exists a subtlety about the prefactor in \eqref{gl1}. Its linear dependence on $\Dl$ which we ignore looks like the anomaly contribution because a scaling dimension is a half of the R-charge $\Dl$. Although this part arises simply from rewriting the Gamma functions and can be absorbed in the normalization factor, now we are not sure that this is really related to $c$-anomaly.

As a future work, we will continue to investigate TBH$_{3}$/qSCFT$_{2}$. We naively expect that the gravity background dual to qSCFT$_{2}$ is the Bandos-Teitelboim-Zanelli  (BTZ) black hole \cite{Banados:1992wn} with some charge \cite{Clement:1993kc, Martinez:1999qi}. In other words, the entropy of the charged BTZ black hole as a solution in three-dimensional supergravity \cite{Izquierdo:1994jz} might be independent of $q$ encoded as the periodicity of the Euclidean time direction, or we might see this expectation quantitatively in the framework of the supergravity embedded into the string theory as discussed in \cite{Giveon:2015cgs}. We would like to understand more physically the $q$-independence of the SRE from the gravitational point of view.

%but we are wondering how we extract explicitly the value of the SRE on the branched sphere. We work on two-dimensional theories, hence, the SRE (or the entanglement entropy) should be proportional to $\log$ with the UV cutoff $\ve$\footnote{$\ve$ is truly a physical one, not identical to the geometrical parameter $\e$ to smooth conical singularities.} as shown in \cite{Giveon:2015cgs} (for general two-dimensional theories, see \cite{Calabrese:2004eu}). We still do not know precise treatment of $\ve$ on the branched sphere. We hope that this will be able to be solved, and the result will be in agreement with general circumstance.

%%%%%%%%%%%%%%%%%% Acknowledgments %%%%%%%%%%%%%%%%%%%%%%%%%%%%
\acknowledgments{The author is grateful to Akinori Tanaka for collaboration at an early stage of the project and beneficial comments. The author also would like to thank Heng-Yu Chen, Kazuo Hosomichi, Bruno Le Floch, Tatsuma Nishioka, and Satoshi Yamaguchi for useful discussions. The work of H.M. was supported in part by the JSPS Research Fellowship for Young Scientists.}

\appendix
%%%%%%%%%%%%%%%%%% section A %%%%%%%%%%%%%%%%%%%%%%%%%%%%%%%%
\section{Preliminaries} \label{Pre}

\paragraph{Convention.} %%%%%%%%%%%%%%%%%%%%%%%%%%%%%%%%%%%%%%%%%%%%
We use the gamma matrices $\g_{a}$ ($a = 1, 2$) and the chirality matrix $\g_{3}$ defined by
\begin{align}
\g_{a} = \sg^{a}, \hspace{2em}
\g_{3} = - i \g_{1} \g_{2} = \sg^{3},
\end{align}
where $\sg^{A}$ ($A = 1, 2, 3$) are Pauli matrices, and $a, b$ are indices of the local Lorentz flame. In addition, we take charge conjugation as $C = \sg^{2}$ with $C^{-1} = C$. This matrix acts on the gamma matrices such that
\begin{align} % cahrge conjugate of gamma matrices
C \g_{A} C^{- 1} &= - \g_{A}^{\rm T}.
\end{align}
We define the inner product of spinors using $C$ as
\begin{align}
\e \la := \e^{\alpha} C_{\alpha \beta} \la^{\beta} = \la \e.
\end{align}
We should note that the Fierz identity for the fermionic spinors\footnote{For the bosonic spinors,
\begin{align}
\xi \lp \e^{\dagger} \la \rp = \frac{1}{2} \lb \la \lp \e^{\dagger} \xi \rp + \g_{\mu} \la \lp \e^{\dagger} \g^{\mu} \xi \rp + \g_{3} \la \lp \e^{\dagger} \g_{3} \xi \rp \rb.
\end{align}}
is given by
\begin{align} \label{Fierz} % Fierz identity
\xi \lp \bar{\e} \la \rp
=
- \frac{1}{2} \lb \la \lp \bar{\e} \xi \rp + \g_{\mu} \la \lp \bar{\e} \g^{\mu} \xi \rp + \g_{3} \la \lp \bar{\e} \g_{3} \xi \rp \rb.
\end{align}
%since it plays an important role to show the closure of supersymmetry.
When $\xi = \la$ and they are matrix-valued, we can obtain the following relations:
\begin{align}
	\begin{aligned}
	0 &= [ \bar{\e} \la, \la ] - [ \bar{\e} \g_{3} \la, \g_{3} \la ] - [ \bar{\e} \g^{\mu} \la, \g_{\mu} \la ], \\
	0 &= \lp \bar{\la} \e_{2} \rp \lp \bar{\la} \e_{1} \rp - \lp \bar{\la} \g_{3} \e_{2} \rp \lp \bar{\la} \g_{3} \e_{1} \rp + \lp \bar{\la} \g_{\mu} \e_{1} \rp \lp \bar{\la} \g^{\mu} \e_{2} \rp.
	\end{aligned} \label{Fierz2}
\end{align}
For our Killing spinors \eqref{ksb2}, the significant bilinear of them is
\begin{align}
\bxi \xi &= - 1, \hspace{2em}
\bxi \g^{A} \xi = \begin{pmatrix} 0 & \sin \theta & \cos \theta \end{pmatrix}. %1
%\xi \xi &= 0, \hspace{2.8em}
%\xi \g^{A} \xi = e^{- i \tau} \begin{pmatrix} - i & - \cos \theta & \sin \theta \end{pmatrix}, \\ %2
%\bxi \bxi &= 0, \hspace{2.8em}
%\bxi \g^{A} \bxi = e^{i \tau} \begin{pmatrix} - i & \cos \theta & - \sin \theta \end{pmatrix}. %3
\end{align}

\paragraph{The Hurwitz zeta function.} %%%%%%%%%%%%%%%%%%%%%%%%%%%%%%%%%%%%%
The Hurwitz zeta function $\zeta ( z, p )$ is defined by
\begin{align} % the Hurwitz zeta function
\zeta ( z, p )
=
\sum_{n = 0}^{\infty} \frac{1}{\lp p + n \rp^{z}}
\label{hurwitz}
\end{align}
for $z, p \in \mathbb{C}$ and $\text{Re} [ z ] > 1$. The derivative of this function with respect to $z$,
\begin{align} % derivative of the Hurwitz zeta function
\frac{\p}{\p z} \zeta ( z, p )
= - \sum_{n = 0}^{\infty} \frac{\log \lp p + n \rp}{\lp p + n \rp^{z}},
\end{align}
has an useful formula
\begin{align}
\log \G ( p )
= \frac{\p}{\p z} \zeta ( z, p ) \Big|_{z = 0} - \frac{\rm d}{{\rm d} z} \zeta ( z ) \Big|_{z = 0},
\label{hzetareg}
\end{align}
where $\G ( p )$ is the Gamma function, and $\zeta ( z )$ is the standard zeta function. We apply \eqref{hzetareg} to regulaitng the divergent product so that
\begin{align} % formula to the Gamma function
\prod_{n \geq 0} \lp p + n \rp
\to
\frac{\sqrt{2 \pi}}{\G ( p )}.
\label{hregformula}
\end{align}

\paragraph{The Gamma function.} %%%%%%%%%%%%%%%%%%%%%%%%%%%%%%%%%%%%%%%
%The definition of the Gamma function is
%\begin{align} % def of the Gamma function
%\G ( z )
%=
%\int_{0}^{\infty} \de t t^{z - 1} e^{- t},
%\end{align}
%and this satisfies the famous difference equations,
The Gamma function satisfies the famous difference equations,
\begin{align} % difference eq.
\G ( z + 1 ) = z \G ( z ), \hspace{2em}
\G ( 1 - z ) = - z \G ( - z ),
\label{pndifference}
\end{align}
where the second relation can be concluded by the important reflection formula
\begin{align} % reflection formula
\G ( z ) \G ( 1 - z )
=
- z \G ( z ) \G ( - z )
=
\frac{\pi}{\sin \pi z}.
\end{align}
Another expression of the Gamma function is Euler's infinite product formula
\begin{align} % Euler's infinite product formula
\G ( z )
=
\frac{1}{z}
\prod_{n = 1}^{\infty}
\lp 1 + \frac{1}{n} \rp^{z}
\lp 1 + \frac{z}{n} \rp^{- 1},
\end{align}
which is used to show the multiplication theorem of Gauss and Legendre \cite{Whittaker}
\begin{align} % the multiplication theorem of Gauss and Legendre
\G ( n z )
=
( 2 \pi )^{\frac{1 - n}{2}}
n^{n z - \frac{1}{2}}
\prod_{k = 0}^{n - 1} \G \lp z + \frac{k}{n} \rp.
\label{multiGL}
\end{align}

%\newpage
%%%%%%%%%%%%%%%%%% section B %%%%%%%%%%%%%%%%%%%%%%%%%%%%%%%%
\section{Localization} \label{Detail}

%%%%%%%%%%%%%%%%%% section B.1 %%%%%%%%%%%%%%%%%%%%%%%%%%%%%%%
\subsection{Supersymmetry algebra} \label{Salg}
As a consistency check for our supersymmetry to be well-defined on $S_{\e}^{2}$, the supersymmetry  algebra for the vector multiplet closes as,
\begin{align} % SUSY algebra for vector
	\begin{aligned}
	\lb \dlz, \dlbz \rb A_{\mu} &= ( \Lcal_{v}^{A} A )_{\mu} + \Dcal_{\mu} \La, \\ %1
	\lb \dlz, \dlbz \rb \rho &= \Lcal_{v}^{A} \rho + i [ \La, \rho ], \\ %2
	\lb \dlz, \dlbz \rb \sg &= \Lcal_{v}^{A} \sg + i [ \La, \sg ], \\ %3
	\lb \dlz, \dlbz \rb \la &= \Lcal_{v}^{A} \la + i [ \La, \la ] - i R_{V} \la, \\ %4
	\lb \dlz, \dlbz \rb \bla &= \Lcal_{v}^{A} \bla + i [ \La, \bla ] + i R_{V} \bla, \\ %5
	\lb \dlz, \dlbz \rb D &= \Lcal_{v}^{A} D + i [ \La, D ], %6
	\end{aligned} \label{alvec}
\end{align}
and for the chiral multiplet closes as,
\begin{align} % SUSY algebra for chiral
	\begin{aligned}
	\lb \dlz, \dlbz \rb \phi &= \Lcal_{v}^{A} \phi + i \La \phi + i \Dl R_{V} \phi, \\ %1
	\lb \dlz, \dlbz \rb \bphi &= \Lcal_{v}^{A} \bphi - i \bphi \La - i \Dl R_{V} \bphi, \\ %2
	\lb \dlz, \dlbz \rb \psi &= \Lcal_{v}^{A} \psi + i \La \psi + i \lp \Dl - 1 \rp R_{V} \psi, \\ %3
	\lb \dlz, \dlbz \rb \bpsi &= \Lcal_{v}^{A} \bpsi - i \bpsi \La - i \lp \Dl - 1 \rp R_{V} \bpsi, \\ %4
	\lb \dlz, \dlbz \rb F &= \Lcal_{v}^{A} F + i \La F + i \lp \Dl - 2 \rp R_{V} F, \\ %5
	\lb \dlz, \dlbz \rb \bF &= \Lcal_{v}^{A} \bF - i \bF \La - i \lp \Dl - 2 \rp R_{V} \bF, %6
	\end{aligned} \label{almat}
\end{align}
where we set the parameters corresponding to the symmetries,
\begin{align}
	\begin{aligned}
	\mbox{translation} &: \hspace{.4em} v_{\mu} &&\hspace{-.9em}= i \bxi \g_{\mu} \xi, \\ %1
	\mbox{gauge transformation} &: \hspace{.45em} \La &&\hspace{-.9em}= \bxi \xi \sg - i \bxi \g_{3} \xi \rho, \\ %2
	%\mbox{dilation} &: \rho &&\hspace{-.9em}= \frac{i}{2} \lp \Dcal_{\mu} \bzeta \g^{\mu} \xi + 	\bzeta \g^{\mu} \Dcal_{\mu} \xi \rp = \frac{1}{2} \Dcal_{\mu} \xi^{\mu}, \\ %3
	\text{R-rotation} &: R_{V} &&\hspace{-.9em}= - \frac{1}{4} \lp \Dcal_{\mu} \bxi \g^{\mu} \xi - \bxi \g^{\mu} \Dcal_{\mu} \xi \rp. %4
	%\mbox{axial R-rotation} &: \beta &&\hspace{-.9em}= \frac{1}{4} \lp \Dcal_{\mu} \bzeta \g_{3} \g^{\mu} \xi - \bzeta \g_{3} \g^{\mu} \Dcal_{\mu} \xi \rp, %5
	\end{aligned}
\end{align}
In the above algebra, $\Lcal_{v}^{A}$ is the gauge-covariant Lie derivative along the $v^{\mu}$ acting on the fields in the adjoint representation as
\begin{align}
	\begin{aligned}
	\Lcal_{v}^{A} A
	&= v^{\mu} F_{\mu \nu} \de x^{\nu}, \\ %1
	\Lcal_{v}^{A} \sg &= v^{\mu} \lp \p_{\mu} \sg - i [ A_{\mu}, \sg] \rp, \\ %2
	\Lcal_{v}^{A} \la &= v^{\mu} \lp \nabla_{\mu} \la - i [ A_{\mu}, \la] \rp + \frac{1}{4} \lp \nabla_{\mu} v_{\nu} \rp \g^{\mu \nu} \la, %3
	\end{aligned}
\end{align}
and on these in the representation $\mathbf{R}$ as
\begin{align}
	\begin{aligned}
	\Lcal_{v}^{A} \phi &= v^{\mu} \lp \p_{\mu} - i A_{\mu} \rp \phi, \\ %1
	\Lcal_{v}^{A} \psi &= v^{\mu} \lp \nabla_{\mu} - i A_{\mu} \rp \psi + \frac{1}{4} \lp \nabla_{\mu} v_{\nu} \rp \g^{\mu \nu} \psi. %2
	\end{aligned}
\end{align}

%%%%%%%%%%%%%%%%%% section B.2 %%%%%%%%%%%%%%%%%%%%%%%%%%%%%%%
\subsection{Vector multiplets} \label{Vector}
We can acquire the linearized SYM Lagrangian by expanding $\Lcal_{\text{SYM}}$ with fluctuations around the locus \eqref{locusvec2}. Certainly, it contains only quadratic terms, and we write the field $\varphi$ in the adjoint representation in terms of the Cartan-Weyl basis
\begin{align}
\varphi = \sum_{i \in {\rm Cartan}} \varphi^{i} H_{i} + \sum_{\alpha \in {\rm root}} \varphi^{\alpha} E_{\alpha},
\end{align}
where $H_{i}$ are Cartan generators, and we use the nomalization $\Tr \lb E_{\alpha} E_{\beta} \rb = \dl_{\alpha + \beta}$. Then, from the linearized Lagrangian, we extract the differential operator $\Dl_{\rm vec}^{b}$ acting on the bosonic fluctuations $( A', \sg', \rho' )^{\rm T}$ in the matrix form as
\begin{align}
\Dl_{\rm vec}^{b} =
	\begin{pmatrix}
	- \ast \Dcal^{(0)} \ast \Dcal^{(0)} + \alpha ( a )^{2} + \frac{\alpha ( s )^{2}}{\ell^{2} q^{2}} %11
	& i \alpha ( a ) \Dcal^{(0)} %12
	& - i \frac{\alpha ( s )}{\ell q} \Dcal^{(0)} + \ast \Dcal^{(0)} \frac{\sqrt{f_{\e}}}{\ell} \\ %13
	i \alpha ( a ) \ast \Dcal^{(0)} \ast %21
	& - \ast \Dcal^{(0)} \ast \Dcal^{(0)} + \frac{\alpha ( s )^{2}}{\ell ^{2} q^{2}} %22
	& \frac{\alpha ( s )}{\ell q} \alpha ( a ) \\ %23
	- i \frac{\alpha ( s )}{\ell q} \ast \Dcal^{(0)} \ast - \frac{\sqrt{f_{\e}}}{\ell} \ast \Dcal^{(0)} %31
	& \frac{\alpha ( s )}{\ell q} \alpha ( a ) %32
	& - \ast \Dcal^{(0)} \ast \Dcal^{(0)} + \alpha ( a )^{2} + \frac{f_{\e}}{\ell^{2}} %33
	\end{pmatrix}, \label{opb1}
\end{align}
where $\ast$ is the Hodge dual, and $\Dcal^{(0)}$ means the covariant derivative defined with the locus value. For later computation, we omit the prime representing the fluctuations.

\paragraph{Gauge fixing.} %%%%%%%%%%%%%%%%%%%%%%%%%%%%%%%%%%%%%%%%%%%
We actually need fixing gauge to obtain the correct one-loop determinant for the vector multiplet. For this purpose, it can be found that the operator $\Dl_{\rm vec}^{b}$ has the following unphysical modes satisfying its eigenvalue equation:
% unphysical modes
\begin{itemize}
\item Zero modes: % zero modes
\begin{align}
\begin{pmatrix} A^{\alpha} \\ \sg^{\alpha} \\ \rho^{\alpha} \end{pmatrix}
=
\begin{pmatrix} i \Dcal^{(0)} \rho \\ - \alpha ( a ) \rho \\ \frac{\alpha ( s )}{\ell q} \rho \end{pmatrix}
\text{ with eigenvalue }
0.
\end{align}

\item Longitudinal modes: % longitudinal modes
\begin{align}
\begin{pmatrix} A^{\alpha} \\ \sg^{\alpha} \\ \rho^{\alpha} \end{pmatrix}
=
\begin{pmatrix} - i \alpha ( a ) \Dcal^{(0)} \rho \\ \tilde{\Dl}_{\rm vec}^{(0)} \rho \\ \frac{\alpha ( s )}{\ell q} \alpha ( a ) \rho \end{pmatrix}
\text{ with eigenvalue }
\lp \tilde{\Dl}_{\rm vec}^{(0)} + \alpha ( a )^{2} \rp,
\end{align}
where $\tilde{\Dl}_{\rm vec}^{(0)}$ is the eigenvalue of the operator
\begin{align}
\Dl_{\rm vec}^{(0)} := - \ast \Dcal^{(0)} \ast \Dcal^{(0)} + \frac{\alpha ( s )^{2}}{\ell^{2} q^{2}}.
\end{align}

\end{itemize}
The zero modes correspond to the gauge symmetry whose gauge volume in the path integral should be removed by introducing the Faddeev-Popov determinant. We follow the short-cut for fixing this gauge explained in \cite{Hama:2011ea}. We should insert the factor $\prod_{i} \dl ( x_{i} )$ in the integration measure to exclude the zero modes, where $x$ represents a set of the eigenmodes. The Faddeev-Popov derteminant $\Dl_{\rm PF}$ is taken as the Jacobian for the change of the variables from $x_{i}$ to the zero mode $\rho$, which can be determined by
%The zero modes correspond to the gauge symmetry whose gauge volume in the path integral should be removed by introducing Faddeev-Popov determinant. We follow the short-cut for fixing this gauge explained in \cite{Hama:2011ea}. The integration measure is written by
%\begin{align} % integration measure
%\Dcal A \Dcal \sg \Dcal \rho
%=
%\prod_{i, \Dl_{\rm vec}^{b} = 0} {\rm d} x_{i} \times \prod_{j, \Dl_{\rm vec}^{b} \neq 0} {\rm d} x'_{j},
%\end{align}
%where $x$ represents a set of the eigenmodes. We should insert the factor $\prod_{i} \dl ( x_{i} )$ to exclude the zero modes. Furthermore, the correct gauge fixing produces the Faddeev-Popov derteminant $\Dl_{\rm PF}$ such that
\begin{align} % Faddeev-Popov derteminant
1
&=
\int \Dcal A \Dcal \sg \Dcal \rho
\exp \lb
- \frac{1}{2} \left. \int \Tr \lp
A^{- \alpha} \w \ast A^{\alpha}
+ \sg^{- \alpha} \w \ast \sg^{\alpha}
+ \rho^{- \alpha} \w \ast \rho^{\alpha}
\rp \right|_{\rm zero\ modes}
\rb \non \\ %1
&=
\Dl_{\rm PF}
\int \Dcal' \rho
\exp \lb \frac{1}{2}
\int \Tr \lp
\rho \w \ast
\lp - \ast \Dcal^{(0)} \ast \Dcal^{(0)}
+ \alpha ( a )^{2}
+ \frac{\alpha ( s )^{2}}{\ell^{2} q^{2}} \rp
\rho \rp \rb, %2
\end{align}
where $\Dcal'$ means the integration excluding the zero modes. In fact, $\Dl_{\rm PF}$ exactly cancels the one-loop contribution from the longitudinal modes \cite{Hama:2011ea}.

% physical modes
The remaining problem to calculate the one-loop determinant for the vector multiplet is to find the pairing structure, that is, which physical (transverse) bosonic and fermionic eigenmodes can be mapped each other by the generators of supersymmetry, $\xi, \bxi$. To do this, it is useful to take a gauge \cite{Gomis:2012wy}
\begin{align} % Gauge fixing
\ast \Dcal^{(0)} \ast A^{\alpha} = i \frac{\alpha ( s )}{\ell q} \rho^{\alpha}, \hspace{2em}
\sg^{\alpha} = 0,
\label{gf}
\end{align}
and then the differential operator $\Dl_{\rm vec}^{b}$ reduces to
\begin{align}
\Dl_{\rm vec}^{b} =
	\begin{pmatrix}
	- \ast \Dcal^{(0)} \ast \Dcal^{(0)} + \alpha ( a )^{2} + \frac{\alpha ( s )^{2}}{\ell^{2} q^{2}} %11
	& - i \frac{\alpha ( s )}{\ell q} \Dcal^{(0)} + \ast \Dcal^{(0)} \frac{\sqrt{f_{\e}}}{\ell} \\ %12
	- i \frac{\alpha ( s )}{\ell q} \ast \Dcal^{(0)} \ast - \frac{\sqrt{f_{\e}}}{\ell} \ast \Dcal^{(0)} %21
	& - \ast \Dcal^{(0)} \ast \Dcal^{(0)} + \alpha ( a )^{2} + \frac{f_{\e}}{\ell^{2}} %22
	\end{pmatrix}. \label{opb2}
\end{align}
In addition to this gauge, as described in \cite{Gomis:2012wy}, we can make the eigenvalue problem of $\Dl_{\rm vec}^{b}$ simpler by defining the operator $\dl_{\rm vec}^{b}$ satisfying
\begin{align}
\det \Dl_{\rm vec}^{b} = \lp \det \dl_{\rm vec}^{b} \rp^{2},
\end{align}
provided that the eigenvalue of $\dl_{\rm vec}^{b}$ is $- i ( M - \alpha ( a ) )$ with some value $M$. One can easily find $\dl_{\rm vec}^{b}$, and we also introduce the differential operator $\Dl_{\rm vec}^{f}$ acting on the fermionic fluctuations as
\begin{align}
\dl_{\rm vec}^{b}
&=
\begin{pmatrix}
i \alpha ( a ) - i \frac{\alpha ( s )}{\ell q} \ast & %11
- \ast \Dcal^{(0)} \\ %12
\ast \Dcal^{(0)} & %21
i \alpha ( s ) - \frac{\sqrt{f_{\e}}}{\ell} %22
\end{pmatrix}, \\
\Dl_{\rm vec}^{f}
&=
i \g^{\mu} \Dcal_{\mu}^{(0)} - i \alpha ( a ) + \frac{1}{\ell q} \g^{3} \alpha ( s ).
\end{align}
Accordingly, what we should do is to discover $M$ by solving the differential equations derived from the action of the operator $\dl_{\rm vec}^{b}$.

\paragraph{Pairing structure.} %%%%%%%%%%%%%%%%%%%%%%%%%%%%%%%%%%%%%%%%%
As the next step, we need to find the boson and fermion eigenmodes which can be mapped each other. For the fermion eigenmodes $\Sg$, provided that
\begin{align}
& \Sg = \lp \g^{\mu} \Acal_{\mu} + \g^{3} \Upsilon \rp \xi, \label{vecbpair21} \\ %1
& \dl_{\rm vec}^{b} \Bcal = - i M \Bcal, \label{vecbpair22} %\hspace{2em}
%\Bcal = \begin{pmatrix} \Acal \\ \Upsilon \end{pmatrix}, %2
\end{align}
where $\Bcal = ( \Acal \ \Upsilon )^{\text{T}}$, we can find the eigenvalue $M$ such that
\begin{align}
\Dl_{\rm vec}^{f} \Sg = i M \Sg.
\label{vecfeigenv}
\end{align}
Similarly for the boson eigenmodes $\Bcal$, when we assume  \eqref{vecfeigenv} and make the map
\begin{align}
\begin{pmatrix}
\Acal \\
\Upsilon \end{pmatrix}
=
\begin{pmatrix}
- i M' \bxi \g_{\mu} e^{\mu} \Sg + i \Dcal^{(0)} \lp \bxi \Sg \rp \\
- i M' \bxi \g^{3} \Sg + \frac{\alpha ( s )}{\ell q} \bxi \Sg
\end{pmatrix},
\label{vecbpair4}
\end{align}
where $M' := M + \alpha ( a )$, we can derive the eigenvalue equation
\begin{align}
\dl_{\rm vec}^{b} \Bcal
= - i M \Bcal.
\label{vecbeigenv}
\end{align}
Thus, the pair of the eigenmodes \eqref{vecbpair21} and \eqref{vecbpair4} gives the trivial one-loop determinant.

\paragraph{Unpaired eigenmodes.} %%%%%%%%%%%%%%%%%%%%%%%%%%%%%%%%%%%%%%%
To obtain the nontrivial contribution, we would like to find the unpaired eigenmodes annihilated by the above maps. We take the ansatz for the unpaired fermion and boson eigenmodes from the vanishing conditions of $\Upsilon$ and $\Sg$, respectively, namely,
\begin{align}
\Upsilon = 0 \text{ in \eqref{vecbpair4}}
&\RA
\Sg
=
g ( \vartheta, \tau ) \lp \g^{3} \bxi - i \frac{\alpha ( s )}{M' \ell q} \bxi \rp, \label{vecfunpair8} \\[.5em] %1
\Sg = 0 \text{ in \eqref{vecbpair21}}
&\RA
\begin{pmatrix}
\Acal \\
\Upsilon
\end{pmatrix}
=
\begin{pmatrix}
\tilde{g} ( \vartheta, \tau ) \lp e^{1} - i \cos \vartheta e^{2} \rp \\
i \tilde{g} ( \vartheta, \tau ) \sin \vartheta
\end{pmatrix}, \label{unpairB1} %2
\end{align}
where $g ( \vartheta, \tau )$ and $\tilde{g} ( \theta, \tau )$ are functions with R-charge $- 2$ and $0$, respectively. This $\Sg$ \eqref{vecfunpair8} must satisfy $\Acal = 0$ and the eigenvalue equation \eqref{vecfeigenv} simultaneously. These conditions seem to be in the overdetermined system because there is only one unknown function $g ( \theta, \tau)$. However, one can find that the solution of $\Acal = 0$ under the assumption \eqref{vecfunpair8} automatically satisfy the eigenvalue equation for $\Sg$. The similar situation happens on \eqref{unpairB1}: one can check that these satisfy the eigenvalue equation \eqref{vecbpair22} and the gauge-fixing condition \eqref{gf} simultaneously in the way that one condition automatically leads to others. 

Then, the condition $\Acal = 0$ and the substitution of \eqref{unpairB1} into the upper component of \eqref{vecbpair22} reduce to the differential equations for $g ( \vt, \tau )$ and $\tilde{g} ( \vt, \tau )$:
\begin{align} % differential equations for g and \tilde{g}
\Acal = 0 \text{ with \eqref{vecfunpair8}}
&\RA
\lc \begin{aligned}
    0
    &=
    i \lp M' + \frac{i \sqrt{f_{\e}}}{\ell} \rp \cos \vartheta g
    + \frac{\alpha ( s )}{\ell q} g
    - \frac{\sqrt{f_{\e}}}{\ell} \sin \vartheta \p_{\vartheta} g, \\ %
    0
    &=
    \lp M' - i \frac{\alpha ( s )}{\ell q} \kappa + \frac{i}{\ell q} \rp g
    + \frac{1}{\ell q} \p_{\tau} g, %
    \end{aligned} \right. \label{vecfunpair4} \\ %1
\begin{aligned} \text{Upper component of \eqref{vecbpair22}} \\ \text{with \eqref{unpairB1}} \end{aligned}
&\RA
\lc \begin{aligned}
	0
	&= \lp M' - i \frac{\sqrt{f_{\e}}}{\ell} \rp \cos \vartheta \tilde{g} - i \frac{\alpha ( s )}{\ell q} \tilde{g} - i \frac{\sqrt{f_{\e}}}{\ell} \sin \vartheta \p_{\vartheta} \tilde{g}, \\ %
	0
	&= i \lp M' - i \frac{\alpha ( s )}{\ell q} \kappa \rp \tilde{g} + \frac{i}{\ell q} \p_{\tau} \tilde{g}. %
	\end{aligned} \right. \label{vecbunpair2} %2
\end{align}
Because these equations in \eqref{vecfunpair4} and \eqref{vecbunpair2} are completely separated in terms of the coordinates, we now can set $g ( \vartheta, \tau) = e^{i j \tau} h ( \vartheta )$ and $\tilde{g} ( \vartheta, \tau) = e^{i j \tau} \tilde{h} ( \vartheta )$ with $j \in \Zbb$. Considering the regular conditions of $h ( \vt )$ and $\tilde{h} ( \vt )$ around the north pole ($\vt \sim 0$) and the south pole ($\vt \sim \pi$) constraints the allowed values for $j$, and the corresponding eigenvalues $M$ are given by
\begin{align} % eigenvalues
\eqref{vecfunpair4} \text{ for fermions}
&\RA
M \ell q 
=
\lc \begin{aligned}
	& - i \lp j - i \alpha ( a ) \ell q + | \alpha ( s ) | \rp && \mbox{for } \alpha ( s ) \neq 0, \\
	& - i \lp j + 1 - i \alpha ( a ) \ell q \rp && \mbox{for } \alpha ( s ) = 0,
	\end{aligned} \right. \label{vecfeigenv2} \\[.5em] %1
\eqref{vecbunpair2} \text{ for bosons}
&\RA
M \ell q
=
- i \lp j - i \alpha ( a ) \ell q - | \alpha ( s ) | \rp, \label{vecbeigenv2} %2
\end{align}
where $j \geq 0$ in \eqref{vecfeigenv2} and $j \leq - 1$ in \eqref{vecbeigenv2}. Note that we do not take the limit $\e \to 0$ in the above process, which means the results can be naively regarded as these on $S_{q}^2$.

\paragraph{One-loop determinant.} %%%%%%%%%%%%%%%%%%%%%%%%%%%%%%%%%%%%%%%
Finally, combining \eqref{vecfeigenv2} and \eqref{vecbeigenv2} provides the one-loop determinant for the vector multiplet (up to the sign factor represented by $\simeq$)
\begin{align} % the one-loop determinant for the vector multiplet
\Zcal_{\text{1-loop}}^{\text{vec}}
&=
\frac{\det \Dl_{\rm vec}^{f}}{\det \dl_{\rm vec}^{b}} \non \\ %1
&\simeq
\prod_{\alpha}
\lb
\prod_{\alpha ( s ) \neq 0}
\frac
{\prod_{j \geq 0} \lp j - i \alpha ( a ) \ell q + | \alpha ( s ) | \rp}
{\prod_{j \leq - 1} \lp j - i \alpha ( a ) \ell q - | \alpha ( s ) | \rp}
\rb
\lb
\prod_{\alpha ( s ) = 0}
\frac
{\prod_{j \geq 0} \lp j + 1 - i \alpha ( a ) \ell q \rp}
{\prod_{j \leq - 1} \lp j - i \alpha ( a ) \ell q \rp}
\rb \non \\ %2
%&\simeq
%\prod_{\substack{\alpha \\ \alpha ( s ) \neq 0}} \prod_{j \geq 0}
%\frac
%{j - i \alpha ( a ) \ell q + | \alpha ( s ) |}
%{j + 1 + i \alpha ( a ) \ell q + | \alpha ( s ) |}
%\non \\ %3
&\simeq
\prod_{\substack{ \alpha > 0 \\ \alpha ( s ) \neq 0}}
\lb \lp \alpha ( a ) \ell q \rp^{2} + \alpha ( s )^{2} \rb %4
\end{align}
%up to the sign factor represented by $\simeq$.

%%%%%%%%%%%%%%%%%% section B.3 %%%%%%%%%%%%%%%%%%%%%%%%%%%%%%%
\subsection{Chiral multiplets} \label{Chiral}
As for the vector multiplet, we get the linearized Lagrangian by expanding $\Lcal_{\text{ch}}$ around the locus \eqref{locusmat2} and define the differential operators
\begin{align}
\Dl_{\text{ch}}^{b}
&=
- \Dcal^{(0) \mu} \Dcal_{\mu}^{(0)}
+ w ( a )^{2}
+ \frac{w ( s )^{2}}{\ell^{2} q^{2}}
+ i \frac{( \Dl - 1) \sqrt{f_{\e}}}{\ell} w ( a )
+ \frac{\Dl ( 2 - \Dl ) f_{\e}}{4 \ell^{2}}, \\ %1
\Dl_{\text{ch}}^{f}
&=
- i \g^{\mu} \Dcal_{\mu}^{(0)}
+ i w ( a )
- \frac{1}{\ell q} \g^{3} w ( s )
- \frac{\Dl \sqrt{f_{\e}}}{2 \ell}, %2
\end{align}
acting on the bosonic and fermionic fluctuations, respectively.

\paragraph{Pairing structure.} %%%%%%%%%%%%%%%%%%%%%%%%%%%%%%%%%%%%%%%%%
Again, we can construct the pairing map between boson eigenmodes and fermion eigenmodes of the chiral multiplet. If we assume the following map from boson $\Phi$ to fermion $\Psi$ and the eigenvalue equation for $\Phi$,
\begin{align}
\begin{pmatrix} \Psi_{1} \\ \Psi_{2} \end{pmatrix}
&=
\begin{pmatrix} \xi \Phi \\ i \g^{\mu} \xi \Dcal_{\mu}^{(0)} \Phi + i \xi w ( a ) \Phi + \xi \lp \frac{1}{\ell q} \g^{3} w ( s ) - \frac{\Dl \sqrt{f_{\e}}}{2 \ell} \rp \Phi \end{pmatrix}, \label{bpair1} \\[.5em] %1
\Dl_{\text{ch}}^{b} \Phi
&=
M \lp M - 2 i w ( a ) + \frac{\Dl - 1}{\ell} \rp \Phi, \label{beigenv} %2
\end{align}
with the eigenvalue $M$, then we can obtain first order differential equations for $\Psi_{1, 2}$
\begin{align} \label{diffeqPsi}
	\begin{aligned}
	% Psi_{1}
	\Dl_{\text{ch}}^{f} \Psi_{1}
	&=
	\lp 2 i w ( a ) - \frac{( \Dl - 1 ) \sqrt{f_{\e}}}{\ell} \rp \Psi_{1} - \Psi_{2}, \\ %1
	% Psi_{2}
	\Dl_{\text{ch}}^{f} \Psi_{2}
	&=
	\lc
	- M \lp M - 2 i w ( a ) + \frac{\Dl - 1}{\ell} \rp \xi
	+ \frac{f'_{\e}}{4 \ell^{2} \sin \theta} \xi \g^{3}
	- i \frac{\Dl f'_{\e}}{4 \ell^{2}} \xi \g^{1}
	\rc \Phi. %2
	\end{aligned}
\end{align}
To produce the results on the branched sphere, we now take the limit $\e \to 0$, that is, $f_{\e} \to 1$ (or equivalently, picking up the zeroth order of $\e$). The equation \eqref{diffeqPsi} with the limit results in the simple eigenvalue equation for $\Psi$ in the matrix form
\begin{align} \label{feigenv}
\Dl_{\text{ch}}^{f} \begin{pmatrix} \Psi_{1} \\ \Psi_{2} \end{pmatrix}
=
\begin{pmatrix} 2 i w ( a ) - \frac{\Dl - 1}{\ell} & - 1 \\ - M \lp M - 2 i w ( a ) + \frac{\Dl - 1}{\ell} \rp & 0 \end{pmatrix}
\begin{pmatrix} \Psi_{1} \\ \Psi_{2} \end{pmatrix}.
\end{align}
Similarly, provided that
\begin{align}
\Phi &= \bxi \Psi, \label{fpair1} \\ %1
\Dl_{\text{ch}}^{f} \Psi &= M \Psi, %2
\end{align}
we can derive a differential equation for $\Phi$
\begin{align}
\Dl_{\text{ch}}^{b} \Phi
&=
\lc
M^{2} \bxi
- 2 i M w ( a ) \bxi
+ \frac{( \Dl - 1 ) \sqrt{f_{\e}}}{\ell} M \bxi
\right. \notag \\ %1
&\hspace{2em} \left.
- i \frac{( \Dl - 1 ) f'_{\e}}{4 \ell^{2}} \bxi \g^{1}
+ \frac{1}{4 \ell^{2}} \cot \theta f'_{\e} \bxi
- \frac{( \Dl - 1 ) f'_{\e}}{4 \ell^{2} \sin \theta} \bxi \g^{3}
\rc \Psi, %2
\end{align}
which still mixes with $\Psi$. Actually, the limit $\e \to 0$ reduces this to the eigenvalue equation
\begin{align} %\label{beigenv2}
\Dl_{\text{ch}}^{b} \Phi
=
M \lp M - 2 i w ( a ) + \frac{\Dl - 1}{\ell} \rp \Phi.
\end{align}

\paragraph{Unpaired eigenmodes.} %%%%%%%%%%%%%%%%%%%%%%%%%%%%%%%%%%%%%%%
Let us turn to find the unpaired modes which cannot be mapped under \eqref{bpair1} and \eqref{fpair1}. For the unpaired fermion modes, we take an ansatz
\begin{align}
\Phi = 0 \text{ in \eqref{fpair1}}
&\RA
\Psi = \bxi g ( \vartheta, \tau ),
\label{unpairchf1}
\end{align}
where $g ( \vartheta, \tau )$ is a function with R-charge $\Dl - 2$. For the unpaired boson modes, when two fermion eigenmodes satisfy the relation
\begin{align}
\Psi_{2} = M \Psi_{1},
\label{unpairchb1}
\end{align}
there does not exist the map from the boson eigenmode $\Phi$ to these fermion eigenmodes. This is because
\begin{align}
\Dl_{\rm ch}^{f}
\begin{pmatrix}
\Psi_{1} \\
\Psi_{2}
\end{pmatrix}
&= 
- \lp M - 2 i w ( a ) + \frac{\Dl - 1}{\ell} \rp
\begin{pmatrix}
\Psi_{1} \\
\Psi_{2}
\end{pmatrix}, %2
\end{align}
that is, the eigenvalue of the fermion eigenmodes is $- \lp M - 2 i w ( a ) + \frac{\Dl - 1}{\ell} \rp$ in this case, and they are not independent each other via \eqref{unpairchb1}. Thus, the eigenvalue $- M$ which does not make a pair with that of the boson eigenmode can contribute to the one-loop determinant.

Substituting \eqref{unpairchf1} and \eqref{unpairchb1} into the eigenvalue equation \eqref{feigenv} and the pairing map \eqref{bpair1}, respectively, reduces to the differential equations for the unpaired modes
\begin{align} % differential equations for g and \Phi
\text{\eqref{feigenv} with \eqref{unpairchf1}}
&\RA
\lc \begin{aligned}
    0
    &=
    \lp M
    - i w ( a )
    + \frac{( \Dl - 2 ) \sqrt{f_{\e}}}{2 \ell} \rp \cos \vartheta g
    + \frac{w ( s )}{\ell q} g
    - \frac{\sqrt{f_{\e}}}{\ell} \sin \vartheta \p_{\vartheta} g, \\ %
    0
    &=
    - i \lp M
    - i w ( a )
    + \frac{w ( s )}{\ell q} \kappa
    + \frac{\Dl - 2}{2 \ell q} \rp g
    + \frac{1}{\ell q} \p_{\tau} g, %
    \end{aligned} \right. \label{chfunpair4} \\ %1
\text{\eqref{bpair1} with \eqref{unpairchb1}}
&\RA
\lc \begin{aligned}
	0
	&=
	- \lp M
	- i w ( a )
	+ \frac{\Dl \sqrt{f_{\e}}}{2 \ell}
	\rp \cos \vartheta \Phi	
	- \frac{w ( s )}{\ell q} \Phi
	- \frac{\sqrt{f_{\e}}}{\ell} \sin \vartheta \p_{\vartheta} \Phi, \\ %
	0
	&=
	\lp M
	- i w ( a )
	+ \frac{w ( s )}{\ell q} \kappa
	+ \frac{\Dl}{2 \ell q}
	\rp \Phi
	+ \frac{i}{\ell q} \p_{\tau} \Phi. %
	\end{aligned} \right. \label{chbunpair2} %2
\end{align}
Obliviously, we can factorize the coordinate dependence of the functions such that $g ( \vartheta, \tau ) = e^{i j \tau} h ( \vartheta )$ and $\Phi ( \vartheta, \tau ) = e^{i j \tau} \tilde{h} ( \vartheta )$ with $j \in \Zbb$. As before, normalizability of $h ( \vt )$ and $\tilde{h} ( \vt )$ imposes restrictions on the possible values of $j$, and then the eigenvalues $M$ for the boson and fermion modes are obtained as
\begin{align} % eigenvalues
\text{\eqref{chfunpair4} for fermions}
&\RA
M \ell q = j + 1 + i w ( a ) \ell q + | w ( s ) | - \frac{\Dl}{2}, \label{feigenv2} \\ %1
\text{\eqref{chbunpair2} for bosons}
&\RA
M \ell q = j + i w ( a ) \ell q - | w ( s ) | - \frac{\Dl}{2}, \label{beigenv2} %2
\end{align}
where $j \geq 0$ in \eqref{feigenv2} and $j \leq 0$ in \eqref{beigenv2}.

\paragraph{One-loop determinant.} %%%%%%%%%%%%%%%%%%%%%%%%%%%%%%%%%%%%%%%
Eventually, combining \eqref{feigenv2} and \eqref{beigenv2} results in the nontrivial one-loop determinant for the chiral multiplet
\begin{align} % the one-loop determinant for the chiral multiplet
\Zcal_{\text{1-loop}}^{\text{ch}}
&=
\frac{\det \Dl_{\text{ch}}^{f}}{\det \Dl_{\text{ch}}^{b}} \non \\ %1
&=
\prod_{w}
\frac
{\prod_{j \geq 0} \lp j + 1 + i w ( a ) \ell q + | w ( s ) | - \frac{\Dl}{2} \rp}
{\prod_{j \leq 0} - \lp j + i w ( a ) \ell q - | w ( s ) | - \frac{\Dl}{2} \rp} \non \\ %2
%&=
%\prod_{w}
%\prod_{j \geq 0}
%\frac
%{j + 1 + i w ( a ) \ell q + | w ( s ) | - \frac{\Dl}{2}}
%{j - i w ( a ) \ell q + | w ( s ) | + \frac{\Dl}{2}} \notag \\ %3
&=
\prod_{w}
\frac
{\G \lp \frac{\Dl}{2} - i w ( a ) \ell q + | w ( s ) | \rp}
{\G \lp 1 - \frac{\Dl}{2} + i w ( a ) \ell q + | w ( s ) | \rp}, %4
\end{align}
where, in the last line, we regularize the infinite products by using the formula of the Hurwitz zeta function \eqref{hzetareg}.

%%%%%%%%%%%%%%%%%% section C %%%%%%%%%%%%%%%%%%%%%%%%%%%%%%%%
\section{Examples of $\Ncal = ( 2, 2 )$ theories}\label{Ex22}
In this appendix, we calculate R-charges $\Dl$ for some 2d $\Ncal = ( 2, 2 )$ theories which flow in the IR to nonlinear sigma models (NLSMs) describing CY manifolds as target spaces. The results confirm the condition $0 < \Dl < 1$, which means that the integer-valued function $[ \eta ]_{\Dl}$ \eqref{cffn} with vorticity \eqref{vorticitych} vanishes in the theories where our exact results are reliable. The program which we apply to obtaining correct R-charges in SCFTs is proposed as $c$-extremization \cite{Benini:2012cz, Benini:2013cda} which is analogue to $a$-maximization in 4d \cite{Intriligator:2003jj}.

\paragraph{Example 1: The quintic.} %%%%%%%%%%%%%%%%%%%%%%%%%%%%%%%%%%%%%%
\begin{align}
	\begin{tabular}{|c||c|c|} \hline
	& U$(1)$ & U$(1)_{R}$ \\ \hline \hline %1
	$\Phi_{a}$ & $+ 1$ & $\Dl_{\Phi}$ \\ %2
	$P$ & $- n$ & $\Dl_{P}$ \\ \hline %3
	\end{tabular}
	\label{glsmP}
\end{align}
The first example is a GLSM describing a CY hyperplane in $\Pbb^{n - 1}$ which contains the fields $\Phi_{a}$ ($a = 1, \cdots, n$) and $P$ shown in \eqref{glsmP}. They are coupled through the superpotential $\Wcal = P f ( \Phi )$ where $f ( \Phi )$ is a polynomial of degree $n$ in $\Phi_{a}$. This superpotential sets the constraint $n \Dl_{\Phi} + \Dl_{P} = 2$, thus, the trial function $\tilde{c}$ is given by
\begin{align}
\frac{\tilde{c}}{3}
&=
n ( \Dl_{\Phi} - 1 )^{2}
+
( \Dl_{P} - 1 )^{2}
-
1 \notag \\ %1
&=
n ( \Dl_{\Phi} - 1 )^{2}
+
( n \Dl_{\Phi} - 1 )^{2}
-
1. %2
\end{align}
Then, the $c$-extremization procedure leads to the R-charges,
\begin{align}
\frac{\de \tilde{c}}{\de \Dl_{\Phi}}
=
0
\RA
\Dl_{\Phi} = \Dl_{P} = \frac{2}{n + 1}
< 1.
\end{align}

\paragraph{Example 2: K3.} %%%%%%%%%%%%%%%%%%%%%%%%%%%%%%%%%%%%%%%%%%
\begin{align}
	\begin{tabular}{|c||c|c|c|} \hline
	& U$(1)_{1}$ & U$(1)_{2}$ & U$(1)_{R}$ \\ \hline \hline %1
	$X_{a}$ & $+ 1$ & $0$ & $\Dl_{X}$ \\ %2
	$Y_{\tilde{a}}$ & $0$ & $+ 1$ & $\Dl_{Y}$ \\ %3
	$P$ & $- 2$ & $- 3$ & $\Dl_{P}$ \\ \hline %4
	\end{tabular}
	\label{glsmK3}
\end{align}
Another simple case is an U$(1)_{1}\times \text{U}(1)_{2}$ gauge theory describing an elliptically fibered K3 as a NLSM at the low energy. There are three kinds of fields $X_{a}$ ($a = 1, 2$), $Y_{\tilde{a}}$ ($\tilde{a} = 1, 2, 3$), and  $P$ which form the superpotential $\Wcal = P f ( X, Y )$ with $f ( X, Y )$ a polynomial of degree $( 2, 3 )$ in $( X_{a}, Y_{\tilde{a}} )$. Accordingly, $\Wcal$ imposes the constraint on these R-charges as $2 \Dl_{X} + 3 \Dl_{Y} + \Dl_{P} = 2$. The trial function $\tilde{c}$ using this constraint is given by
\begin{align}
\frac{\tilde{c}}{3}
&=
2 ( \Dl_{X} - 1 )^{2}
+
3 ( \Dl_{Y} - 1 )^{2}
+
( \Dl_{P} - 1 )^{2}
-
2 \notag \\ %1
&=
2 ( \Dl_{X} - 1 )^{2}
+
3 ( \Dl_{Y} - 1 )^{2}
+
( 2 \Dl_{X} + 3 \Dl_{Y} - 1 )^{2}
-
2, %2
\end{align}
then $c$-extremization provides the R-charges as solutions for simultaneous equations,
\begin{align}
\frac{\p \tilde{c}}{\p \Dl_{X}} = \frac{\p \tilde{c}}{\p \Dl_{Y}} = 0
\RA
\Dl_{X} = \Dl_{Y} = \Dl_{P} = \frac{1}{3}
< 1.
\end{align}

\paragraph{Example 3: The resolved $\mathbb{WP}_{1,1,2,2,2}^{4} [8]$.} %%%%%%%%%%%%%%%%%%%%%
\begin{align}
	\begin{tabular}{|c||c|c|c|} \hline
	& U$(1)_{1}$ & U$(1)_{2}$ & U$(1)_{R}$ \\ \hline \hline %1
	$X_{a}$ & $0$ & $+ 1$ & $\Dl_{X}$ \\ %2
	$Y_{\tilde{a}}$ & $+ 1$ & $0$ & $\Dl_{Y}$ \\ %3
	$Z$ & $+ 1$ & $- 2$ & $\Dl_{Z}$ \\ %4
	$P$ & $- 4$ & $0$ & $\Dl_{P}$ \\ \hline %5
	\end{tabular}
	\label{glsmWP}
\end{align}
There is a simple but nontrivial example of a GLSM which describes a Calabi-Yau threefold (CY3) which is the resolution of a weighted degree $8$ hypersurface in a 4d weighted projective space $\mathbb{WP}_{1,1,2,2,2}^{4} [8]$ \cite{Candelas:1993dm, Morrison:1994fr}. The field contents are summarized in \eqref{glsmWP} ($a = 1, 2$ and $\tilde{a} = 1, 2, 3$). The superpotential is set to be $\Wcal = P f ( X, Y, Z )$ where $f ( X, Y, Z )$ is a wighted homogeneous polynomial of degree $( 2, 3, 1 )$ in $( X_{a}, Y_{\tilde{a}}, Z )$, which gives the constraint $2 \Dl_{X} + 3 \Dl_{Y} + \Dl_{Z} + \Dl_{P} = 2$. The trial function $\tilde{c}$ is written by
\begin{align}
\frac{\tilde{c}}{3}
&=
2 ( \Dl_{X} - 1 )^{2}
+
3 ( \Dl_{Y} - 1 )^{2}
+
( \Dl_{Z} - 1 )^{2}
+
( \Dl_{P} - 1 )^{2}
-
2 \notag \\ %1
&=
2 ( \Dl_{X} - 1 )^{2}
+
3 ( \Dl_{Y} - 1 )^{2}
+
( \Dl_{Z} - 1 )^{2}
+
( 2 \Dl_{X} + 3 \Dl_{Y} + \Dl_{Z} - 1 )^{2}
-
2. %2
\end{align}
The $c$-extremization procedure gives simultaneous equations whose solutions are the precise values of the R-charges,
\begin{align}
\frac{\p \tilde{c}}{\p \Dl_{X}} = \frac{\p \tilde{c}}{\p \Dl_{Y}} = \frac{\p \tilde{c}}{\p \Dl_{Z}} = 0
\RA
\Dl_{X} = \Dl_{Y} = \Dl_{Z} = \Dl_{P} = \frac{2}{7}
< 1.
\end{align}

%\paragraph{Example 4: R{\o}dland's Pfaffian CY3.} %%%%%%%%%%%%%%%%%%%%%%%%%%%%%%%
%\begin{align}
%	\begin{tabular}{|c||c|c|} \hline
%	& U$(2)$ & U$(1)_{R}$ \\ \hline \hline %1
%	$\Phi_{a = 1, \cdots, 7}$ & $\mathbf{1}_{- 2}$ & $2 - 4 \hfrak$ \\ %2
%	$P_{i = 1, \cdots, 7}$ & $\square_{+ 1}$ & $2 \hfrak$ \\ \hline %3
%	\end{tabular}
%	\label{glsmRP}
%\end{align}
%Field contents of a GLSM describing a R{\o}dland's Pfaffian Calabi-Yau threefold. The subscript represents the charge under the determinant of U$(2)$.

%%%%%%%%%%%%%%%%%% References %%%%%%%%%%%%%%%%%%%%%%%%%%%%%%%
%\bibliographystyle{utphys}
%%\nocite{*}
%\bibliography{../2dVortexDefect/2dVortexDefectRef}
\providecommand{\href}[2]{#2}\begingroup\raggedright\endgroup

\end{document}